\documentclass[11pt,a4paper]{article}
\usepackage[margin=1in]{geometry}
\usepackage{amsmath,amssymb,booktabs,graphicx,siunitx}
\usepackage[T1]{fontenc}
\usepackage[dvipsnames]{xcolor}
\usepackage[numbers,sort&compress]{natbib} 
\usepackage{hyperref}
\hypersetup{%
colorlinks=true,
linkcolor=blue!50!black, 
citecolor=blue!50!black, 
urlcolor=blue!50!black,
pdftitle={The Electron-Recoil Response Tail and the LZ High-Energy Event Background},
pdfauthor={Triparno Bandyopadhyay,Rajarajeswari Ramasamy,Nishitha TP},
pdfsubject={Background Uncertainty in LZ result of one anomalous event that is interpreted as Dark Matter},
pdfkeywords={LZ, Dark Matter, Electron Recoil},
pdflang={en},
pdfproducer={LaTeX with hyperref},
pdfcreator={pdflatex}
} 
\usepackage[nameinlink]{cleveref}
\usepackage{orcidlink}
\newcommand{\keVee}{\ensuremath{\mathrm{keV_{ee}}}} 
\usepackage{authblk}

\author[1]{Triparno
Bandyopadhyay\,\orcidlink{0000-0002-6744-8972}\thanks{\href{mailto:gondogolegogol@gmail.com}{\texttt{gondogolegogol@gmail.com}}}}
\author[2]{Rajarajeswari
Ramasamy\,\orcidlink{0009-0004-8452-6751}\thanks{\href{mailto:rajarajeswari.r.work@gmail.com}{\texttt{rajarajeswari.r.work@gmail.com}}}}
\author[2]{Nishitha
TP\,\orcidlink{0009-0005-8208-0976}\thanks{\href{mailto:nishitha.tp2001@gmail.com}{\texttt{nishitha.tp2001@gmail.com}}}}

\affil[1]{Nayanta University, Pune 411021, Maharashtra, India. }
\affil[2]{Department of Physics and Nanotechnology, SRM Institute of
Science and Technology, Kattankulathur, Tamil Nadu 603203, India.}

\title{\bfseries The Electron-Recoil Response Tail and the\\ LZ
High-Energy Event Background} 
\date{\today}

\begin{document}
\maketitle

\begin{abstract}
\noindent
We reconstruct the continuum electron-recoil contribution to the LZ
background expectation from the collaboration's released simulation
configuration. The released model specifies a definite response tail, but
the published calibration information supplies no bound on how far the
physical tail may depart from it at the probabilities the expectation
depends on. Distributions reproducing the executable model's mean,
variance, and standardised skewness exactly, and its event-conditioned
10, 50, and 90\% \(N_e\) quantiles to about one per cent, differ in the
relevant tail probability by more than five orders of magnitude. The
ladder is short: a tail agreeing with that central characterisation to
0.14\% in the quantiles lifts the electron-recoil contribution out of
the negligible range, and one agreeing to 0.47\% makes it comparable to
the entire background quoted for the comparison region. We go on to
identify the measurements that would settle it.
\end{abstract}

\section{Introduction}%
\label{sec:intro}

The LUX-ZEPLIN (LZ) Collaboration~\cite{LZ:2019sgr} has reported one event in a
2.84~tonne-year exposure consistent with a \(248\pm23\,\mathrm{stat}
\pm23\,\mathrm{sys}\)~keV nuclear recoil, at \(\mathrm{S1c}=540.1\)~phd
and \(\mathrm{S2c}=9268\)~phd~\cite{LZ:2026axp}\footnote{The Letter gives
\(\mathrm{S2c}=9268\)~phd in its text and again in the supplementary
discussion of the multiple-scintillation hypothesis, and we use that value
throughout. The caption of its Fig.~4 instead places the event at
\(10^{3.98}\)~phd, which is \(9550\)~phd; \(\log_{10}9268=3.967\). The two
differ by three per cent, which moves \(N_e\) by eight electrons and the
event's own displacement with it, though not the \(\mathrm{S2c}\) ceiling on
which the leakage probability is defined.} (the LZ paper is from
now on referred to as `the Letter')\footnote{ A number of papers have
	since examined possible sources of the
	excess~\cite{Freese:2026mfi,Anwar:2026ffj,Pradhan:2026inq,%
		Su:2026rwz,Freese:2026sga,Wu:2026nhi,Yin:2026jnn,%
		DiMauro:2026ldr,Visinelli:2026kgt,%
		Yamashita:2026ump,Chattopadhyay:2026ryw,Smirnov:2026aqk,Unwin:2026rdp,%
		Du:2026guj,McCabe:2026crm,Jeesun:2026vzo,Rodd:2026tyn,Dent:2026bji,Gu:2026vto,%
		deLima:2026shq,Lee:2026wof,Baer:2026fpy,Wang:2026ytg,Yang:2026wpb,%
		DiMauro:2026dqp,Liang:2026coz,Kotlarski:2026pep,Das:2026uyy,Okada:2026eol,%
		Alhazmi:2026efz,Ahmed:2026qjg,Du:2026lpa,Bandyopadhyay:2026gjw,Borah:2026zwf,%
		Bisal:2026khf,Kannike:2026qyl,Bose:2026ndd,Khan:2026nwp,Yuan:2026djt,%
		Cheung:2026byg,Bamwidhi:2026vdu,Elahi:2026vlm,Zhu:2026dag,%
		Aghaie:2026vsu,Lee:2026jxl,Asadi:2026iot,Lee:2026xxh,Langhoff:2026ujr,%
		Fan:2026hzw,Chattaraj:2026fxn,He:2026hqz,%
		Kumar:2026lgi,Heikinheimo:2026kwp,Nguyen:2026lui,Lee:2026zbr,Okada:2026upm,%
		Mahapatra:2026glu,Borah:2026ris,Barman:2026omh,Le-Yaouanc:2026djt,Ghosh:2026txe,%
		Das:2026buc,Lian:2026hpm,Baer:2026yrt,Palmisano:2026kuj,%
		Uttayarat:2026isp,He:2026idw,DiMauro:2026ymt,Arcadi:2026kev,An:2026pkc,%
		Ge:2026xax,Xing:2026civ,Nagata:2026pbj,Okada:2026fef,%
		OHare:2026nqi,Bose:2026szs,Paul:2026okh,Ahmed:2026com,De:2026win,Sheng:2026tqt,%
		Khan:2026zuj,Khan:2026osp,Sannino:2026hkc,Chauhan:2026udz,Jung:2026otm,%
		Dong:2026mck,Gemmell:2026yaw,Delepine:2026ith,Maier:2026qpr,%
		Khan:2026pgu,Girmohanta:2026qzo,Capolupo:2026uyi,Langhoff:2026qgo,%
		Murayama:2026apt,Kalita:2026hvb,Brdar:2026ukx,%
		CentellesChulia:2026ats,Dutra:2026fng,Pallis:2026jdc,%
		Oliveira:2026jtn,Das:2026jsq,Bozorgnia:2026emp,Chanda:2026wdl,%
		Jho:2026zzj,Jho:2026cso,Meissner:2026ooy,%
Roy:2026wzb,Choudhury:2026qjz,Hor:2026voa}.}. In the
\(500<\mathrm{S1c}<600\)~phd comparison region where the event lies, the
total expected background is quoted as
\(0.0106\pm0.0008\,(\mathrm{sys})\) counts. An expectation of that size
is a statement about the far tail of a response distribution, and this
paper asks what the published information determines about it.

That expectation is what carries the discrimination, because the event's
own observables do not settle it. The Supplement reports that ``the S1
pulse timing has limited discrimination power on an event-by-event basis
for events with \(\mathrm{S1c}\sim550\)~phd and the event of interest is
generally compatible with both ER and NR populations'', that AmBe nuclear
recoils near \(150\)~keV ``are not inconsistent with an ER template built
from \(^{131m}\)Xe calibration data near the position of the event of
interest'', and that its S1 hit pattern agrees ``at the \(2\sigma\) level
in \(z\) and \(1\sigma\) level in \((x,y)\)'' with single-scatter
electron recoils of similar size and
location~\cite{LZ:2026axp}. Separating the two hypotheses therefore rests
on the expected rate rather than on the event's pulse shape or position,
which is why the size of that expectation is the quantity at issue.

Our central result is that the published characterisation of the
response does not determine the probability that matters, and carries no
stated uncertainty on the quantity that does. In \cref{sec:family} we
construct families of distributions that reproduce the executable
model's mean, variance and standardised skewness exactly, and its
event-conditioned \(10\), \(50\), and \(90\%\) \(N_e\) quantiles to about
one per cent, and that differ in the relevant tail probability by more
than five orders of magnitude. Three such families, spanning two
distinct construction mechanisms, agree on this within a factor of about
two at matched quantile deviation. The thresholds that follow are
correspondingly close: agreement to \(0.14\%\) in the quantiles lifts
the electron-recoil (ER) contribution out of the negligible range, and
\(0.47\%\) makes it comparable to the entire background quoted for the
region (\cref{sec:leak}). These families are not offered as competing
physical models of the liquid-xenon response, and the range they span is
not a confidence interval on the leakage. They are constructive
counterexamples: distributions that satisfy the central characterisation
defined in \cref{sec:family}---our summary of the released model, not an
accuracy the collaboration claims---while giving different probabilities
in the extreme tail. What is at issue throughout is therefore what the
published summary of the response determines, not whether the released
model is right, and not the collaboration's internal calibration data;
nor do we carry the result through to the reported significance
(\cref{sec:leak}).

The question is not one the Letter leaves unmentioned. Discussing the
\(^{124}\)Xe and \(^{125}\)I double-vacancy modes, which release
\(64.3\) and \(67.3\)~keV and so fall inside the parent window of
\cref{sec:leak}, it records that these ``are modeled following
Ref.~\cite{Recomb}, and there is a systematic uncertainty in projecting
into the tail of the distribution that we do not include in the
statistical inference''~\cite{LZ:2026axp}. That is a tail-projection
systematic, named and deliberately left out of the fit, for components
at the energy at issue. What this paper adds is the size of the
corresponding effect for the dominant continuum components, where no
such caveat is attached.

The Letter anticipates the obvious objection. Having noted a systematic
in projecting into the tail of the double-vacancy distributions that it
does not include in the statistical inference, it adds that
\begin{quote}
    we note the lack of events between the event of interest and the
    bottom of the ER band in Fig.~4~\cite{LZ:2026axp}.
\end{quote}
That is, if the tail were as heavy as our families permit, intermediate
events should appear, and none are seen. We compute it, and find in
\cref{sec:family} that the interval observed to be empty is narrow
enough that every family predicts well under one event in it, so the
observation does not discriminate between them. 

\Cref{sec:family} carries the central argument of the paper. The
extrapolation of the calibrated ordinary ER response to a probability of
\(\mathcal{O}(10^{-7})\) is treated there and in \cref{sec:leak}.
Implementation divergences that do not change the result materially are
collected in \cref{app:impl}. \Cref{sec:recon} inverts the event into
quanta, reconstructs the executable response model and validates an
analytic representation of it against \(1.1\times10^{10}\) simulated
events, and \cref{sec:settle} states what measurements would determine
the tail, at what precision, and how many calibration events a shape
fit would need if it were attempted in place of a tail measurement.

We work with the released simulation configuration rather than from the
printed specification, and the two differ. Evaluating Eq.~4 of the
supplement of the Letter (`the Supplement' from now on) gives
\(\omega=0.0682\)\footnote{\(\omega\) is the recombination-fluctuation
parameter of the NEST detector response model.}, where the released NEST
implementation~\cite{NESTcode,Szydagis:2022ikv} returns \(0.0462\) at
the same \(N_q\); the ER leakage probability differs between them by a
factor of \(3.2\times10^{3}\). The implementation value is the one
compatible with the published expectation: propagated through the parent
spectrum, the printed form of Eq.~4 gives an electron-recoil
contribution of \(0.60\) counts in the comparison region, far above the
\(0.0106\pm0.0008\) the Letter quotes for \emph{all} backgrounds there.
Taken literally with the printed parameters, and folded through the
parent spectrum and acceptance reconstructed here, Eq.~4 is therefore
incompatible with the published expectation. We therefore instantiate
NEST's \texttt{LZ\_Detector\_2024} class under the \textsf{EFT2024}
configuration specified in the next paragraph, call the yield and quanta
routines on it, and extract the response width from the variance they
return, using the implementation value throughout.

\begin{sloppypar}
All simulation results use \texttt{SetEFT2024Configuration()} in
\texttt{include/\allowbreak Detectors/\allowbreak LZ\_WS2024.hh} of the NEST
repository~\cite{NESTcode,Szydagis:2022ikv} at commit \texttt{541b643}, the
configuration itself having been added in commit \texttt{307fb27} of 27
August 2026. The Letter's main text cites NEST v2.4.5, its Supplement
discusses v2.4.0, and the detector file's own header recommends v2.4.0;
we work from the repository at the commit above rather than from a
tagged release, because the \texttt{EFT2024} parameter vector selects a
branch of \texttt{RecombOmegaER} (the \(\log_{10}N_q\) two-component
form, chosen when the component centre exceeds unity) that the earlier
releases do not contain. Every number below is reproducible at that
commit. What ties that commit to the Letter is circumstantial: the configuration is named for the
analysis, it carries the Letter's detector parameters, and it was added to
the public repository shortly before the preprint appeared. We have not
established that it is the build the collaboration ran,
and if the response differed from this one the quantitative reconstruction
of \cref{sec:leak} would change---though the identifiability result of
\cref{sec:family} would not, since it needs only some definite response to
deform. The driver, scan and analysis scripts are archived at
DOI~\textsc{\href{https://zenodo.org/records/23318901}{10.5281/zenodo.23318901}}~\cite{Archive}, as described in the
code and data availability statement below. 
\end{sloppypar}

Three distinct regions appear throughout the analysis. The \emph{WS RoI}
is the search region of the Letter, \(3<\mathrm{S1c}<600\)~phd,
\(\mathrm{S2}>645\)~phd and \(10^{2.75}<\mathrm{S2c}<10^{4.15}\)~phd. It
is a cut in observables, not in energy. We write simply ``RoI'' for it
where no other region is in play; regions belonging to other analyses
are named explicitly. The \emph{comparison region} is the
\(500<\mathrm{S1c}<600\)~phd subset of the WS RoI shown in the lower
panel of Fig.~5 of the Letter, for which the expectation
\(0.0106\pm0.0008\) counts is quoted; the event lies here. The lowest panel
carries no label of its own in v1 of the Letter: the figure is stated to
cover three \(\mathrm{S1c}\) ranges, the upper two are marked
\(\mathrm{S1c}<250\)~phd and \(250<\mathrm{S1c}<500\)~phd, and the WS
RoI ends at \(600\)~phd, so the lower panel is
\(500<\mathrm{S1c}<600\)~phd. We read it that way throughout.  The
\emph{parent window} is the range of ER parent energies able to satisfy
the comparison region's requirements at all, \(61.3<E<79.1\)~\keVee,
derived in \cref{sec:leak}.

Two derived energies recur which are not cuts. The first is
\(20.02\)~\keVee, the parent energy at which an electron recoil with
\emph{mean} recombination produces \(N_e=409\)~e\(^-\) and therefore
sits exactly at the \(\mathrm{S2c}\) ceiling; parents above it enter the
WS RoI only through charge suppression or a downward fluctuation. The
second is \(0.62\)~\keVee, the lowest parent energy that can satisfy
the three WS RoI cuts jointly. Together they bound the interval over
which the WS RoI accepts ordinary electron recoils, which serves as the
denominator when the Letter's Table~I counts are converted to rates.

We write \(N_q\) for the total quanta produced in a deposit, \(N_i\) for
the number of electron--ion pairs created before recombination,
\(N_{ex}\) for the number of directly excited atoms, \(N_{ph}\) and
\(N_e\) for the photons and escaping electrons finally observed, and
\(r\) for the recombination fraction, and \(\rho_{ex}\equiv N_{ex}/N_i\)
for the exciton-to-ion ratio, so that \(N_q=N_{ph}+N_e\),
\(N_i=N_q/(1+\rho_{ex})\), \(N_e=(1-r)N_i\) and \(N_{ph}=N_{ex}+rN_i\).
\(Q\) denotes the ground-state \(Q\)-value of a \(\beta\) decay, the
maximum kinetic energy shared by the emitted leptons. Finally, \(P\) is
the {per-decay conditional probability} that an electron recoil
of parent energy in the relevant range satisfies the observable criteria
of the comparison region, and \(N^{\mathrm{ER}}_{\mathrm{leak}}\) the
{integrated expectation} obtained by folding \(P\) with the
parent spectrum.

\section{The executable response model}%
\label{sec:recon}

We condition throughout on the observed total quanta rather than on a
nominal energy, because the quanta are what the observables determine,
and because recombination conserves \(N_q\), it moves an event along a
contour of constant total quanta, redistributing photons and electrons
without changing their sum. With \(g_1=\) \SI{0.1104}{phd/photon} and
\(g_2=\) \SI{34.5}{phd/electron} from the released configuration (the
Letter quotes \(g_1=0.110\pm0.002\) phd/photon),
\begin{align}
    \label{eq:e1}
    N_{ph}\;&=\;\frac{\mathrm{S1c}}{g_1}\;=\;4892\,,\quad
    N_e\;=\;\frac{\mathrm{S2c}}{g_2}\;=\;269\,,\quad
    N_q\;=\;N_{ph}+N_e\;=\;5161\;.
\end{align}
With \(W=13.532\)~eV, the mean energy per quantum at these conditions,
the observed quanta correspond to a parent energy
\(E=N_qW=69.84\)~\keVee. The released simulation returns an
exciton-to-ion ratio \(\rho_{ex}=N_{ex}/N_i=0.181\), so the number of
electron--ion pairs created before recombination is
\(N_i=N_q/(1+\rho_{ex})=4369\). The surviving charge is
\(N_e=269\)~e\(^-\), so the recombination fraction implied by the
observation is \(r=1-N_e/N_i=0.939\): \(94\%\) of the ionisation has
recombined. The tuned charge-yield model \(Q_y(E)\), the mean charge
yield in electrons per keV~\cite{Szydagis:2022ikv,Recomb}, predicts a
nominal \(\bar{r}=0.717\) for a typical electron recoil at the same
energy. The event is therefore far more recombined than the model
expects. The deposited energy is unremarkable; under an electron-recoil
hypothesis the distinguishing feature is the partition between photons
and electrons. Note that two observables cannot separately determine the
parent energy \(E\), the recombination fraction \(r\), and a surviving
charge fraction \(f\). Where charge is lost the observed charge is
\(N_e^{\mathrm{obs}}=f(1-r)N_i\) and does not by itself fix \(r\); the
photons are not lost, so \(r\) follows from \(N_{ph}=N_{ex}+rN_i\)
instead. In this paper we work with the \(f=1\) branch throughout. 

We instantiate the collaboration's detector configuration and call
\texttt{GetYields} and \texttt{GetQuanta} directly, so that yields,
widths, density, and the \(W\)-value convention follow the released
configuration. At \(N_q\simeq5.2\times10^{3}\) the model returns
\texttt{recombProb}=0.716 and \texttt{Variance}=\(4.35\times10^{4}\). We
define the recombination-fluctuation width \(\omega\) by
\begin{align}
    \label{eq:e5}
    \omega\;&=\;\frac{\sqrt{\mathrm{\texttt{Variance}}
        -\bar{r}(1-\bar{r})N_i}}{N_i}\;,
\end{align}
the subtracted term being the binomial contribution to the variance.  The
recombination fraction here is \(\bar{r}\), the \texttt{recombProb} the
model returns at that \(N_q\) and not the \(r=0.939\) the event itself
implies; substituting the latter does not reproduce \(\omega\).  It
is the fractional standard deviation of the recombination
probability at fixed \(N_q\), as obtained from the implementation. At
the simulated energy, \(E=70.95~\mathrm{keV_{ee}}\), Eq.~\eqref{eq:e5}
gives \(\sigma_{N_e}=\sqrt{\texttt{Variance}}=208.5\)~e\(^-\). At the
event-conditioned energy, \(E=69.84~\mathrm{keV_{ee}}\)
(\(N_q=5161\)), it gives \(\sigma_{N_e}=204.1\)~e\(^-\), and it is
this value that is used throughout.

The quantity \(\omega\) is not a constant of the model. Both the
returned variance and \(N_i\) depend on \(N_q\), and \(\omega\) with
them: it is \(0.0462\) at the event's \(N_q=5161\) and \(0.0465\) at
\(N_q=5243\). The Supplement writes the same quantity as a function
\(\sigma_p(x)\) of \(x=\log_{10}N_q\); the implementation returns it as
\texttt{RecombOmegaER}. As we shall see, the two specifications give
different values at the same \(N_q\).

\begin{figure}[t]
\centering
\includegraphics[width=\textwidth]{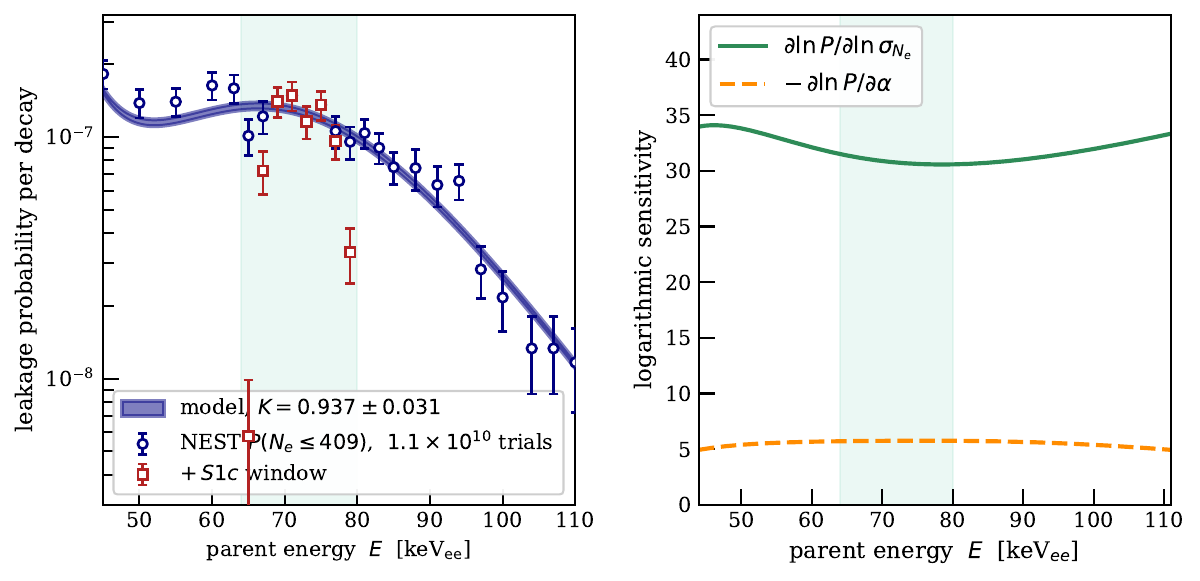}
\caption{Leakage probability per decay (left) and its logarithmic
    sensitivities (right), from \(1.1\times10^{10}\) simulated events
    under the released \textsf{EFT2024} configuration. Circles: \(P(N_e\le409)\),
    the S2c ceiling of the WS RoI\@. Squares: the additional requirement
    \(\mathrm{S1c}\in[500,600]\)~phd. Shading marks the parent energies
    able to satisfy both. The sensitivities are approximately constant
    in energy across the range, so the leverage is a property of the
response model rather than of the event's particular energy.}
\label{fig:f2}
\end{figure}

The executable model samples \(N_e\) from a skew-normal. We construct the
corresponding density and distribution function explicitly, so that the
leakage probability can be evaluated by quadrature rather than by sampling.
The density is elementary; but the distribution function requires
Owen's \(T\) function~\cite{Owen}, which is standard and well conditioned
but has no elementary closed form.
\begin{align}
    \label{eq:skewnormal}
    f(x)=\frac{2}{\omega_\mathrm{sn}}\,\varphi(z)\,\Phi(\alpha z)\,, \qquad
    z=\frac{x-\xi}{\omega_\mathrm{sn}}\,,\qquad
    \varphi(u)=\frac{1}{\sqrt{2\pi}}e^{-u^{2}/2}\,,
\end{align}
with \(\Phi\) the standard normal CDF, location \(\xi\), scale
\(\omega_\mathrm{sn}>0\) and shape \(\alpha\); at \(\alpha=0\) it
reduces to a Gaussian, and it is the \(\kappa=2\) member of the family
introduced later as \cref{eq:family}. Writing \(\mu(E)=Q_y(E)\,E\) for
the mean charge, \(\sigma_{N_e}(E)\) for the width of
\cref{eq:skewnormal}, and taking \(\gamma_\mathrm{ER}(E)\) for its shape
parameter \(\alpha\), the location and scale that reproduces the first two
moments are
\begin{align}
    \label{eq:5a}
    \delta\;&=\;\frac{\alpha}{\sqrt{1+\alpha^{2}}}\,,\qquad    
    \omega_\mathrm{sn}\;=\;\frac{\sigma_{N_e}}{\sqrt{1-2\delta^{2}/\pi}}\,,\qquad
    \xi\;=\;\mu-\omega_\mathrm{sn}\,\delta\sqrt{2/\pi}\,,
\end{align}
and the leakage probability per decay is\footnote{%
We write \(P_\mathrm{win}(E)\) for the same quantity evaluated from the
simulated events of \cref{fig:f2} rather than from the analytic
representation. Their window integrals agree to \(0.2\%\); per-energy
agreement is not established by the scan, whose statistical error is
\(18\%\) per point, and what the scan determines is the single
normalisation \(K\) to \(3.3\%\) together with the deviance of
\cref{eq:e6}.
Where the analytic form is used below it is written \(P(E)\);
where the simulated points are used they are written
\(P_\mathrm{win}(E)\).}
\begin{align}
    \label{eq:5b}
    P(E)\;&=\;K\left[\Phi_{\rm sn}\!\left(N_e^{\rm hi}\right) 
        - \Phi_\mathrm{sn}\!\left(N_e^\mathrm{lo}\right)\right]\,, \qquad
    \Phi_\mathrm{sn}(x)=\Phi(z)-2\,T(z,\alpha),\qquad
    z=\frac{x-\xi}{\omega_\mathrm{sn}}.
\end{align}
\Cref{eq:5a} inverts the standard skew-normal moment relations and \cref{eq:5b}
uses its distribution function, both due to
Azzalini~\cite{Azzalini}. The integration limits carry both RoI
conditions: the S2c ceiling and the S1c window to give
\begin{align}
    N_e^\mathrm{hi}=\min\!\left(409.5,\;N_q-\frac{500}{g_1}\right)\,, \qquad
    N_e^\mathrm{lo}=\max\!\left(0,\;N_q-\frac{600}{g_1}\right)\,.
\end{align}
The upper limit is the S2c ceiling \(10^{4.15}/g_2=409.43\) applied to a
count. The condition admits every integer \(N_e\le409\), and a continuous
density reproduces the probability of that at the half-integer edge, so the
skew-normal distribution function is evaluated at \(409.5\) throughout
and the quantity we write \(P(N_e\le409)\) is that value. The two limits
coincide at both ends of the parent window: at \(E\simeq\)
\(61.3\)~\keVee, they meet at \(N_e=0\) and at \(E\simeq\)
\(79.1\)~\keVee\ they meet at the ceiling. Between these two extremes,
the interval has non-zero width, reaching its maximum at the middle of
the window. At the event-conditioned energy, \cref{eq:5a} gives
\(\delta=0.884\), \(\omega_\mathrm{sn}=288.0\) and
\(\xi=1034.6\)~e\(^-\), and \cref{eq:5b} returns \(P=1.3\times10^{-7}\).

\begin{table}[t]
\begin{center}
\begin{tabular}{p{0.11\textwidth}p{0.34\textwidth}p{0.44\textwidth}}
\toprule
Quantity & Meaning & Where it is specified\\
\midrule
\(\alpha_1,\alpha_2\) & skew parameters of \(\sigma_p(x)\) & Supplement Eq.~4 / Table~S4\\
\(\omega\) & recombination fluctuation & Eq.~4 and implementation differ by 1.475\\
\(\gamma_\mathrm{ER}\) & skew-normal \emph{shape} parameter of the \(N_e\) sampling, which the implementation names a skewness; the standardised skewness it implies is \(\gamma_1=0.4232\) & Implementation (LUX Run04 model)\\
\texttt{OldW13eV} & \(W\)-value convention & Released configuration\\
\(m_6\) & ER-yield sentinel, not coefficient & Table~S3 and configuration differ in sign\\
\bottomrule
\end{tabular}
\end{center}
\caption{Provenance of values and conventions required to perform the calculation leading to \cref{fig:f2}.}
\label{tab:tab1}
\end{table}

The single free quantity is the overall normalisation \(K\) which we fix
by comparison against direct simulation of \(1.1\times10^{10}\) events
at 24 energies spanning 45--\(110\)~\keVee, accumulating 935 leakage
counts. A single overall normalisation describes the whole range. Write
\(p_i\) for the analytic probability at the \(i\)th scan point,
evaluated at unit normalisation from the mean, standard deviation, and
standardised skewness that NEST returned there, and \(T_i\) for the
number of trials. The leakage counts are Poisson,
\(n_i\sim\mathrm{Pois}(K\,p_iT_i)\), so the maximum-likelihood estimate
of \(K\) is closed-form, \(\hat{K}=\sum_i n_i\big/\sum_i p_iT_i\). It
gives 
\begin{align} 
	\label{eq:e6} K\;&=\;0.937\pm0.031\,,\qquad
D/\mathrm{dof}\;=\;0.9\quad(20.8/23,\;p=0.59)\;, 
\end{align} 
with \(D\) the Poisson deviance, which is the appropriate statistic for
counts this small. All 24 points enter; none has zero counts against the
ceiling alone, which is the column the fit uses. The column that
additionally imposes the S1c window does return zero at the lowest
energies, and that is a different criterion: it is the one quoted in
\cref{sec:leak}. Pooled over the 24 points the fit is exact by
construction, \(K\) being the one quantity adjusted; what it establishes
is that a single constant suffices. The scatter of the measured
probabilities about the analytic curve is \(18\%\) rms against a mean
statistical error of \(18\%\) per point, so no residual energy
dependence survives, and the normalisation itself is determined to
\(3.3\%\) (\cref{fig:f2}). The analytic representation therefore
reproduces the simulated leakage scan across the range, the relevant
few-\(\sigma\) tail included, with a single normalisation constant fixed
to a few per cent. Residual structure at the edges of the S1c acceptance
window is attributable to \(N_q\) (Fano) smearing, which the analytic
representation omits. The integrals of \cref{sec:leak} use the simulated
points directly and are unaffected. Reproducing the calculation requires 
the values of five quantities and a convention. We document each in 
\cref{tab:tab1}. 

Equation~4 of the Supplement gives \(\sigma_p(x)\), \(x=\log_{10}N_q\), as a sum of two
skew-Gaussians,
\begin{align}
    \label{eq:sigmap}
    \sigma_p(x)\!&=\!\sum_{i=1,2}A_i\,
    e^{-(x-\mu_i)^2/2\sigma_i^2}
    \left[1+\mathrm{erf}\!\left(\frac{\alpha_i(x-\mu_i)}{\sqrt{2}\,\sigma_i}\right)\right],
\end{align}
with amplitudes \(A_i\), centres \(\mu_i\), widths \(\sigma_i\) and skew
parameters \(\alpha_i\) taken
from Table~S4 of the Supplement: \(A_1=0.03174\), \(\mu_1=2.588\), \(\sigma_1=0.3627\),
\(\alpha_1=-0.3583\) and
\(A_2=0.06211\), \(\mu_2=5.200\), \(\sigma_2=1.359\),
\(\alpha_2=-2.599\).
Evaluated at the event-conditioned \(N_q=5161\), the expression returns
\(\omega=0.06817\), where the released implementation
(\texttt{RecombOmegaER}) returns \(0.04622\) and direct extraction via
\cref{eq:e5} confirms the same value to three figures, \(0.0462\). The
ratio of printed \(\omega\) to implemented is therefore \(1.4750\), and
it is reproduced by the component-wise
mode normalisation carried in the implementation, which divides
component \(i\) by its own value at its mode \(x_i^\mathrm{mode}\).
Both values are evaluated at the event-conditioned \(N_q=5161\). Because
the two component normalisations are constants, the ratio is very nearly
independent of \(N_q\): it is \(1.4744\) at \(N_q=4529\) and
\(1.4751\) at \(5435\), the ends of the parent window, and remains
within half a per cent of \(1.475\) over
\(3\times10^{3}<N_q<8\times10^{3}\). At \(N_q=5243\) the printed form
gives \(0.0685\) and the implementation \(0.0465\).

The two values of \(\omega\) give \(\sigma_{N_e}=299.3\)~e\(^-\) and
\(204.1\)~e\(^-\) at the event-conditioned energy, and hence leakage
probabilities of \(4.27\times10^{-4}\) and \(1.32\times10^{-7}\), a
factor of \(3.2\times10^{3}\). The ratio of the charge widths,
\(1.4664\), is slightly below the ratio of the \(\omega\) values,
\(1.4750\), because the binomial term in \cref{eq:e5} is common to the
two and dilutes the difference; the leakage ratio is not,
because \(P\) is exponential in \(\sigma_{N_e}\) at this displacement,
and a forty-eight per cent increase in the width becomes the factor of
\(3.2\times10^{3}\) above. Extrapolating from one leakage probability to the
other with the local logarithmic derivative would overstate this
substantially, since that derivative falls steeply across the interval.

The published and implemented forms are therefore not on an equal
footing, and we must choose between them. Folded through the parent
spectrum and acceptance of \cref{sec:leak}, the published form of Eq.~4
of the Supplement would give an electron-recoil contribution of \(0.60\)
counts in the comparison region, far above the total of
\(0.0106\) counts the Letter quotes for all components. Hence, we use the
implementation value throughout. A second consideration points the same
way. The Supplement introduces the two-component form of Eq.~4 precisely
because the single skewed Gaussian of NEST v2.4.0 ``was found not to
reproduce the measured ER-band widths across the full calibration energy
range'', and records that the replacement ``provides good agreement with
the measured ER-band widths across the calibrated range''. The two
readings of Eq.~4 differ by \(1.475\) in \(\omega\) and so cannot both
reproduce those widths; the claim of agreement is made for whichever of
them the collaboration evaluated. The argument is one-directional and we do
not stretch it: the printed form is \emph{excluded} by the published total,
while the implementation value is merely \emph{compatible} with it, as any
value giving an electron-recoil contribution below about \(0.01\) counts
would be. That admits nearly two decades of tail probability around the
value we compute, and nothing published lets us confirm that the
collaboration's own electron-recoil component equals our
\(1.8\times10^{-4}\). The \(0.60\)-count figure should not be
read as a claim about the internal LZ analysis; it establishes only that
Eq.~4, taken literally with the printed parameters, is not by itself the
complete specification of the response that produced the published
background. 

Table~S4 of the Letter publishes \(\alpha_1=-0.3583\) and \(\alpha_2=-2.599\). These are
the skew parameters of the \emph{function} \(\sigma_p(x)\), not of the
\(N_e\) distribution. The executable model applies a separate shape
parameter, \(\gamma_\mathrm{ER}\), to the \(N_e\) sampling, taken from the
LUX Run04 skewness model of Sec.~2.2.3 of Ref.~\cite{Szydagis:2022ikv};
at the event-conditioned energy it evaluates to,
\(\gamma_\mathrm{ER}=1.892\). The implementation calls this a skewness and
we keep its name, but it is the skew-normal shape parameter of
\cref{eq:5a} and not a standardised skewness: the standardised skewness
of \(N_e\) that it implies is \(\gamma_1=0.4232\), and it is that
quantity, not \(\gamma_\mathrm{ER}\), which the families below reproduce
exactly. That model is not switched on by default in later versions of the
code, so its being active here is a property of the configuration rather
than of NEST, and we establish it from the scan: the standardised skewness
returned at the twenty-four scan energies is nowhere near zero and agrees
with \(\gamma_1(\gamma_\mathrm{ER}(E))\) to about one per cent, which
also fixes the convention, a standardised skewness being bounded by
\(0.9953\). We use the evaluated value
rather than the rounded \(1.9\) throughout, so that the family
construction, the figures and the leakage integrals all rest on the same
response. Running it through
the \texttt{SkewnessER} argument of \texttt{GetQuanta} at fixed width, the
leakage spans a factor of 57 over  \(\gamma_\mathrm{ER}\in[1.55,2.25]\),
with a local derivative of \(\partial\ln
P/\partial\gamma_\mathrm{ER}=-5.8\) at the default value. 

\section{The published characterisation does not determine the tail}%
\label{sec:family}

In this section we establish two things. The first is quantitative: how
\emph{small} a departure from the central characterisation is already
consequential for this inference, which is the ladder of \cref{sec:leak}
and is not something one can guess. The second is that no uncertainty on
the tail shape accompanies the published response at all, where a
comparable characterisation elsewhere in the field supplies one
(\cref{sec:settle}). The families are the instrument for establishing
the first, not the claim.

The expected background in the comparison region rests on a per-decay
probability \(P\simeq1.3\times10^{-7}\). Writing \(\mu\) for the mean of
the \(N_e\) distribution and \(\sigma_{N_e}\) for its standard
deviation, the S2c ceiling lies at a displacement
\begin{align}
    \frac{\Delta}{\sigma_{N_e}} = \frac{\mu - 409.5}{\sigma_{N_e}}\;,
\end{align}
from the mean. At the event-conditioned energy this is \(4.1\sigma\).

The displacement should be read against the range over which the
response model was tuned. The caption of Fig.~2 of the
Letter~\cite{LZ:2026axp} identifies its grey points as those ``outside
of \(4\sigma\) from the mean of the ER and NR bands'' and states that
``these points are not used in deriving the \textsc{nest} model'',
attributing them to interactions near the walls. The model is therefore
fitted to calibration data inside \(4\sigma\), and the background
expectation for the comparison region is evaluated just outside it. The
Letter separately places the event itself \(1.5\sigma\) below the median of
the modelled NR band at its \(\mathrm{S1c}\) and \(6.7\sigma\) below the median
of the modelled ER band---a different measure from the \(4.1\sigma\)
above, and not to be reconciled with it. That figure is the event's own
displacement from the ER band median at fixed \(\mathrm{S1c}\), in units
of the band width in \(\log_{10}\mathrm{S2c}\), where the band mixes
parent energies, whereas \(4.1\sigma\) is the displacement of the
\emph{ceiling} from the mean of \(N_e\) at fixed \(N_q\). Only the first is
commensurate with the \(4\sigma\) of the calibration cut, which is stated in
that same band coordinate; the \(4.1\sigma\) is a measure of where our
leakage integral is evaluated within our own conditional distribution, and we
do not offer it as evidence of lying outside the calibrated region. We draw no inference from the
exclusion of those points, which is a sensible precaution against wall
leakage and is not a claim about the tail. The point is narrower still: the
probability the background expectation rests on is reached by extrapolating
the calibrated response to \(\mathcal{O}(10^{-7})\), and the published
information examined here does not quantify the uncertainty of that
extrapolation. That is the circumstance in which the question of this
section arises, and it does not require the two displacements to be
mapped onto one another.

The question this section asks is what the published characterisation of
that response determines about the probability at such a displacement.
We address it by constructing families of distributions that reproduce the
central characterisation defined below and examining their behaviour in
the tail. These families are not proposed as physical alternatives for the
liquid-xenon response, and the range they exhibit is not a confidence
interval or a model-averaged uncertainty. They are counterexamples to
identifiability: explicit distributions that satisfy that benchmark while
producing substantially different probabilities in the relevant extreme
tail. Which tail shapes are
physically realised is a question that additional response information
would settle, and one we do not address.

The executable model samples \(N_e\) from a skew-normal. We embed that
in a one-parameter family by replacing the Gaussian kernel of the skewed
construction with a generalised normal~\cite{Subbotin,Nadarajah} of shape \(\kappa\):

\begin{align}
    p_\kappa(x)\;&=\;\frac{2}{s}\,g_\kappa(z)\,G_\kappa(\alpha z)\,,\qquad 
    z\;=\;\frac{x-\xi}{s}\,,\qquad 
    g_\kappa(u)\;=\;\frac{\kappa}{2\,\Gamma(1/\kappa)\,\kappa^{1/\kappa}}\,e^{-|u|^\kappa/\kappa}\,,
\label{eq:family}
\end{align}
with \(G_\kappa\) the CDF of \(g_\kappa\). The factor 2 follows from the
symmetry of \(g_\kappa\), so \(p_\kappa\) is normalised for every
\(\kappa\); we use the \(\exp(-|u|^\kappa/\kappa)\) convention, other
conventions differ by a rescaling of \(s\). At \(\kappa=2\) the scale
\(s\) coincides with \(\omega_{\rm sn}\) of \cref{eq:skewnormal}. At
\(\kappa=2\) one has \(g_2=\varphi\) and \(G_2=\Phi\), so
\cref{eq:family} is exactly \(2\varphi(z)\Phi(\alpha z)/s\), the
skew-normal the executable model samples. The symmetric kernel carries
no skewness of its own: the asymmetry comes entirely from
\(G_\kappa(\alpha z)\).

For each \(\kappa\) the three parameters \((\xi,s,\alpha)\) are obtained
by solving the three equations that set the first moment, the second
central moment, and the standardised third central moment of
\(p_\kappa\). The solution is unique over the range tested. Scanning
\(\alpha\in[0.01,10]\) on a 400-point grid at every \(\kappa\) used returns
exactly one root of \(\gamma_1(\alpha)=0.4232\), at
\(\alpha=\{2.43,2.25,1.89,1.52,1.33,1.13,0.70\}\) for the seven values
\(\kappa=\{2.3,2.2,2.0,1.8,1.7,1.6,1.4\}\) of the table below. The third moment is
monotonic in \(\alpha\) through the root, though for \(\kappa>2\) it dips
negative as \(\alpha\to0\), far below the target. The matched
skewness follows from \(\gamma_\mathrm{ER}\) by
\begin{align}
\delta=\frac{\gamma_{\rm ER}}{\sqrt{1+\gamma_{\rm ER}^{2}}}=0.8841,
\qquad
\gamma_1=\frac{4-\pi}{2}\,
\frac{\left(\delta\sqrt{2/\pi}\right)^{3}}{\left(1-2\delta^{2}/\pi\right)^{3/2}}=0.4232 ,
\label{eq:skewconv}
\end{align}
the two being the same distribution in different
parameterisations~\cite{Azzalini}. The generalised normal is chosen
because it contains the executable model as a member; it is not
privileged.

For each \(\kappa\), the three moment-matching conditions give a
distribution with the same mean, variance, and standardised skewness as
the executable model. What distinguishes the members of the family is
the shape of the tail. We characterise each member by two quantities:
the largest fractional deviation of its \(10\), \(50\), and \(90\%\)
quantiles from those of the executable model, and the resulting tail
probability \(P(N_e\le409)\), which is what determines the background
expectation. \Cref{tab:family} gives both, together with the ratio of the
tail probability to the executable model's and the corresponding
electron-recoil contribution, for a range of deviations from \(0.16\%\)
to \(1.08\%\). The heavier-tail branch, \(\kappa<2\), is tabulated in
more detail than the lighter, \(\kappa>2\), because the question at
issue is whether the background could be larger than the executable
model predicts, not smaller. 

\begin{table}
\begin{center} 
\begin{tabular}{crrrrrr}
\toprule 
\(\kappa\) & q.\ shift & dens.\ dev. & \(P(N_e\le409)\) & \(R\) & \(N^\mathrm{ER}_\mathrm{leak}\) & \(R_{\rm ev}\)\\ 
\midrule 
2.3 & 0.22\% & 4.3\%  & \(4.24\times10^{-10}\) & 0.0032 & \(5.9\times10^{-7}\) & \(7.3\times10^{-5}\)\\ 
2.2 & 0.16\% & 3.0\%  & \(4.06\times10^{-9}\) & 0.0308 & \(5.7\times10^{-6}\) & \(3.5\times10^{-3}\)\\ 
2.0 & ---    & ---    & \(1.32\times10^{-7}\) & 1 & \(1.8\times10^{-4}\) & 1\\ 
1.8 & 0.23\% & 3.9\%  & \(1.72\times10^{-6}\) & 13 & \(2.4\times10^{-3}\) & 49\\ 
1.7 & 0.39\% & 6.5\%  & \(4.95\times10^{-6}\) & 37 & \(6.9\times10^{-3}\) & 220\\ 
1.6 & 0.59\% & 9.7\%  & \(1.28\times10^{-5}\) & 97 & \(1.8\times10^{-2}\) & 814\\ 
1.4 & 1.08\% & 18.3\% & \(6.88\times10^{-5}\) & 521 & \(9.6\times10^{-2}\) & \(7.0\times10^{3}\)\\ 
\bottomrule 
\end{tabular} 
\end{center}
\caption{Members of the one-parameter family of Eq.~\eqref{eq:family}.
  Each row is the member whose generalised-normal shape parameter is
  \(\kappa\); the row \(\kappa=2\) is the executable model itself. The
  second column is the largest fractional shift in the \(10\), \(50\),
  and \(90\%\) quantiles of \(N_e\) from those of the executable model,
  \(\max_p |x_p - x_p^{\rm NEST}| / x_p^{\rm NEST}\). The third is the
  largest fractional deviation in the probability density over the
  same interval. The fourth is the tail probability
  \(P(N_e\le409)\) at the event-conditioned energy; the fifth is the
  ratio \(R\) of that probability to the executable model's; the sixth
  is the corresponding electron-recoil contribution
  \(N^{\rm ER}_{\rm leak} = R \times 1.84\times10^{-4}\) counts, the
  baseline obtained in Sec.~\ref{sec:leak}. The final column, \(R_{\rm
  ev}\), is the ratio of probability densities at the event's own
  charge, \(N_e = 268.6\)~e\(^-\), which is the quantity an unbinned
  likelihood weighs.}
\label{tab:family}
\end{table}

\begin{figure}[t]
\centering
\includegraphics[width=0.74\textwidth]{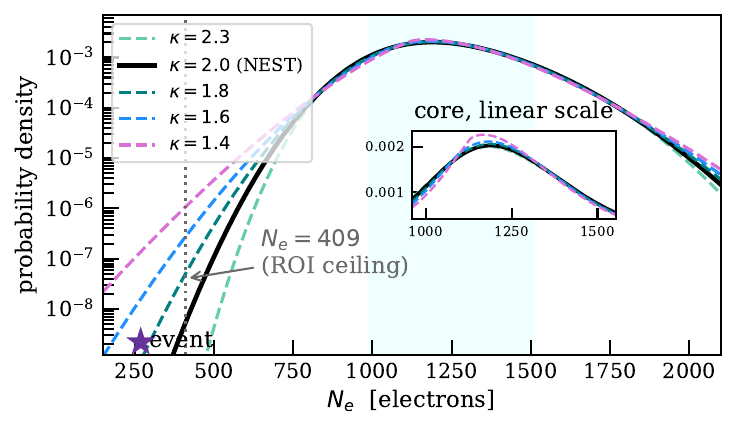}
\caption{Members of the family of \cref{eq:family} with mean, standard
    deviation and standardised skewness fixed at the executable model's
    values. Through the shaded \(10\)--\(90\%\) NEST interval at
    \(E=\)69.84 \keVee\  the curves are not
    resolved by the central characterisation (quantiles agree
    to \(\sim1\%\); the inset shows the densities on a linear scale). Below
\(N_e\simeq500\) their probability densities separate by about three
decades. The dotted line is the S2c ceiling \(N_e=409\); the star marks
the event.}
\label{fig:family}
\end{figure}

\noindent The second column is the largest fractional \emph{shift} in the
10, 50, and 90\% quantiles of \(N_e\) relative to \(\kappa=2\), that is,
\(\max_p|x_p-x_p^{\mathrm{NEST}}|/x_p^{\mathrm{NEST}}\) over
\(p=0.10,0.50,0.90\). It is a displacement in electron number at fixed
probability, not an integrated probability over an interval. These are the quantiles of \(N_e\) at the event-conditioned energy,
the displacement at which the leakage probability is itself defined. They are
not the quantiles of the band drawn in Fig.~2 of the Letter, which is
built at fixed \(\mathrm{S1c}\) from a flat energy spectrum; the
quantity here characterises the response model, not that band. We refer
to the executable model's first three moments together with these three
quantiles as its \emph{central characterisation}, and all deviations
quoted are measured against it. The term is ours and so is the choice of
constraints: this is our summary of the released model, not an accuracy
the collaboration claims. The only stated accuracy we are aware of is
the \(0.2\%\) in mean \(\log_{10}\mathrm{S2c}\) of \cref{sec:settle},
which refers to the centre of the response over the WIMP search range
and not to its tail at this energy. A distribution reproducing the
central characterisation to a quarter of a percent changes the leakage
by a factor of \(13\); one reproducing it to about one percent changes
it by \(521\). The last column, \(R_{\rm ev}\), is the ratio of the
probability \emph{density} at the event's own charge,
\(N_e=268.6\)~e\(^-\), rather than of the probability integrated up to
the ceiling; it is the quantity an unbinned likelihood weighs, and
because the event sits further into the tail than the ceiling does it is
steeper by roughly an order of magnitude, reaching \(7.0\times10^{3}\)
at \(\kappa=1.4\) where the integrated ratio is \(521\). We quote the
integrated ratio throughout, as the conservative of the two. At
\(\kappa=1.66\) (solving continuously in \(\kappa\) rather than
interpolating the table) the electron-recoil contribution alone equals
the total background quoted for the region. \cref{fig:family} shows the
degeneracy directly.

The third column gives the largest fractional deviation in the
probability \emph{density} over the same interval. Within each branch
both deviations increase monotonically with \(|\kappa-2|\), but the
heavier-tail branch deviates more than the lighter one at equal
\(|\kappa-2|\). Both the density and quantile maxima occur at the
\(10\)\textsuperscript{th} percentile, the lowest of the three quantiles, where the tails begin to
separate, and the deviation is much smaller through the bulk. A shape
fit to high-statistics calibration data could discriminate among these
families at the level of the density, whereas the published summary of
the response does not.

The tail-probability ratios \(P(N_e\le409)/P_{\kappa=2}(N_e\le409)\) listed
in the table are computed at the event-conditioned energy. If instead both
numerator and denominator are integrated over the parent window
\(61.3 < E < 79.1\)~\keVee, with the parent rate flat across it
as in \cref{sec:leak} and each family member rebuilt at every energy to the
moments the executable model has there, the resulting ratios change by at most
two percent: for \(\kappa = 1.8, 1.7, 1.6, 1.4\) the window-weighted values
are \(12.8\), \(36.9\), \(96.3\), \(521\) against single-energy \(13\),
\(37\), \(97\), \(521\). The single-energy values are therefore used
throughout, and the approximation costs far less than the factors of
\(10\)--\(10^{3}\) on which the argument turns. 

Our findings are a property of the constraint set rather than of the
family used to exhibit it, and we test this in two stages. First we keep
the construction of \cref{eq:family} and change only the kernel,
replacing \(g_\kappa,G_\kappa\) by the density and CDF of a
Student-\(t\) with \(\nu\) degrees of freedom,
\begin{align}
\label{eq:family2}
    p_\nu(x)\;=\;\frac{2}{s}\,t_\nu(z)\,T_\nu(\alpha z),\qquad z=\frac{x-\xi}{s},
\end{align}
which has polynomial rather than exponential-power tails and recovers
the skew-normal as \(\nu\to\infty\), with \((\xi,s,\alpha)\) fixed by
the same three moment conditions.

Second, we change the construction itself. The Johnson \(S_U\) family is
obtained by a transformation rather than by skewing a
kernel~\cite{Johnson}. If \(Z\) is standard normal then
\(X=\xi+\lambda\sinh[(Z-\gamma)/\delta]\). The random variable \(X\)
here is the electron number \(N_e\). So
\begin{align}
\label{eq:family3}
    F(x)\;=\;\Phi\!\left[\gamma+\delta\,\mathrm{asinh}\!\left(\frac{x-\xi}{\lambda}\right)\right],
\end{align}
with \((\gamma,\xi,\lambda)\) fixed by the same three moment conditions
and \(\delta\) left free. Note that \(\xi\) and \(\delta\) here are not
the skew-normal parameters of \cref{eq:5a}; the two families are
independent constructions. Unlike the other two families it does not
contain the skew-normal as a member, so it tests whether the result
survives when the executable model is not even available as a limit.

\begin{figure}[t]
\centering
\includegraphics[width=0.72\textwidth]{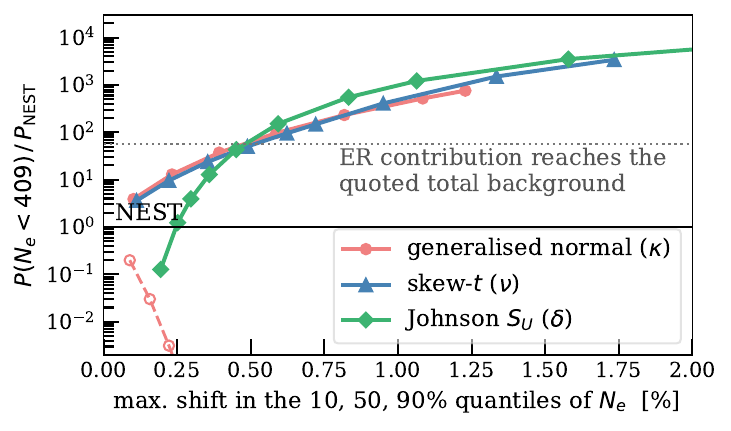}
\caption{Tail probability relative to the executable model, against the
	largest fractional shift in the \(10\), \(50\), and \(90\%\)
	quantiles of \(N_e\), for the two families of \cref{eq:family}
	and~\cref{eq:family2} and the transformation family of
	\cref{eq:family3}. Marker shape identifies the family: circles
	for the generalised normal, triangles for the skew-\(t\),
	diamonds for the Johnson \(S_U\). The generalised normal is
	drawn as two branches, \(\kappa<2\) with filled circles and a
	solid line and \(\kappa>2\) with open circles and a dashed line;
	the other two are each a single solid curve, the skew-\(t\)
	lying wholly above the executable model and the Johnson \(S_U\)
	crossing it. The two kernel families lie close to each other
	over most of the range, agreeing to \(3\%\) at \(0.8\%\)
	deviation and diverging to \(36\%\) by \(1.2\%\); the Johnson
	\(S_U\) family, which does not contain the skew-normal, spans
	both sides of it and rises more steeply, remaining within a
	factor of about two of the others at matched deviation. The
	dotted line marks where the electron-recoil contribution of
	\cref{sec:leak} would equal the total background quoted for the
	comparison region, reached at a quantile deviation of about half
a percent.} 
\label{fig:ratio} 
\end{figure}

\Cref{fig:ratio} plots the tail probability against agreement with the
quantile shift defined above for all three families: the quantile
deviation is the quantity the central characterisation pins down, and the
ratio is what the inference depends on. The two kernel families lie
close to each other over most of the range, agreeing to \(3\%\) at
\(0.8\%\) deviation and diverging to \(36\%\) by \(1.2\%\), where both
are already far above the executable model. The Johnson \(S_U\) family rises more steeply and
sits above them, by a factor of \(1.8\) at \(0.6\%\) deviation and
\(2.4\) at \(1.0\%\):

To test whether the growth of the ratio, \(R\), with the quantile
deviation is a property of the construction or of the constraint set, we
compare the three families at matched deviation. For each family we
solve for the parameter value that gives a specified shift in the 10,
50, and 90\% quantiles of \(N_e\) and evaluate the corresponding tail
ratio; the rows in \cref{tab:families} give that comparison at four 
deviations spanning sub-percent agreement. 

\begin{table}
\begin{center} 
\begin{tabular}{crrrr} 
\toprule
quantile deviation & generalised normal & skew-\(t\) & Johnson \(S_U\) & spread\\ 
\midrule 
\(0.6\%\) & 102 & 86 & 158 & \(\times1.8\)\\ 
\(0.8\%\) & 218 & 211 & 464 & \(\times2.2\)\\
\(1.0\%\) & 404 & 485 & 980 & \(\times2.4\)\\ 
\(1.2\%\) & 706 & 957 & 1613 & \(\times2.3\)\\ 
\bottomrule 
\end{tabular} 
\end{center}
\caption{Tail-probability ratios for the three families of
  Eqs.~\eqref{eq:family}, \eqref{eq:family2} and \eqref{eq:family3},
  compared at matched agreement with the central characterisation.
  Each row corresponds to a specified shift in the \(10\), \(50\), and
  \(90\%\) quantiles of \(N_e\); within a row, each family's shape
  parameter is solved so that its largest fractional quantile shift
  from the executable model equals the value in the first column. The
  three middle columns give \(P(N_e \le 409)/P_{\rm NEST}\), the tail
  probability of the family member relative to the executable model at
  the event-conditioned energy. The final column is the ratio of the
  largest to the smallest of the three entries in that row. The rows
  are chosen to span sub-percent agreement; the corresponding values
  are plotted in Fig.~\ref{fig:ratio}.}
\label{tab:families}
\end{table}

\noindent The entries in the middle three columns are tail probabilities
relative to the executable model, \(P(N_e\le409)/P_{\rm NEST}\),
evaluated at the quantile deviation indicated. The final column gives
the ratio of the largest to the smallest entry in each row.

Three families spanning two construction mechanisms therefore
align within a factor of about two at matched agreement with the
quantile shift, while all give ratios of order \(10^2\) at sub-percent
agreement. Across the three families tested, the relationship between
quantile agreement and tail ratio is approximately family-independent,
even though the tail ratio itself varies by three orders of magnitude
over the range scanned. That the correspondence survives the change of
functional form points to the moment constraints, rather than the choice
of form, as what sets it; the residual factor-of-two spread is itself a
measure of what the published characterisation leaves open. An increase
of \(16.6\%\) in \(\sigma_{N_e}\) brings the electron-recoil
contribution to the total quoted for the region; a change in tail shape
reaching the same place alters those quantiles by about \(0.5\%\). The 
tail probability is therefore determined by the shape of the response 
at the relevant displacement to a much greater degree than by its width.

The empirical objection to this line of reasoning, quoted in
\cref{sec:intro}, is: if the tail were as heavy as those families
permit, events should appear between the event of interest and the
bottom of the ER band, and none are observed. Let us quantify what that
observation constrains. Within \(500<\mathrm{S1c}<600\)~phd the observed
population of Fig.~4 extends down to \(\log_{10}\mathrm{S2c}=4.30\),
that is \(N_e=578\)~e\(^-\); the lowest point nearby, at
\(\log_{10}\mathrm{S2c}=4.25\), lies at \(\mathrm{S1c}=495\)~phd and so
outside the region. The interval observed to be empty is therefore
\(269<N_e<578\), between the event and the lower edge of the population.
Folding the parent spectrum of \cref{sec:leak} through it, we get the
figures in \cref{tab:gap}.

As the table shows, zero events is the expected outcome for 
every family considered, including those whose tail probability exceeds 
the nominal one by a factor of several hundred. The heaviest tail in the 
table would leave the interval empty half the time; at four times the 
present exposure a null result there would disfavour it at the few-per-cent
level. The one input taken from Fig.~4 of the Letter is the lower edge of 
the population at \(\log_{10}\mathrm{S2c}=4.30\); reading it at \(4.35\) or 
\(4.40\) instead raises the heaviest family's expectation to \(1.6\) or \(3.5\)
counts, at which point a null observation would begin to tell against
it, while reading it at \(4.25\) lowers it to \(0.33\). The conclusion that
no family is excluded holds for any edge at or below \(4.35\). The
observation is consistent with the nominal response and with every
alternative we construct, and does not discriminate between them.

The reason is one of scale rather than of selection. The same
calculation gives \(2.3\times10^{3}\) electron recoils in this
\(\mathrm{S1c}\) slice across the band as a whole, which is why the
population there is a continuum; the interval at issue lies several
standard deviations below that bulk while remaining far above the
displacement at which the leakage probability is defined. An observation
over such an interval separates only models differing by orders of
magnitude more than these do.

\begin{table}
\begin{center}
\begin{tabular}{crr}
\toprule
\(\kappa\) & expected counts in \(269<N_e<578\) & \(P(\text{observe }0)\)\\
\midrule
2.0 (NEST) & 0.021 & 0.98\\
1.8 & 0.081 & 0.92\\
1.7 & 0.145 & 0.86\\
1.6 & 0.250 & 0.78\\
1.5 & 0.420 & 0.66\\
1.4 & 0.689 & 0.50\\
\bottomrule
\end{tabular}
\end{center}
\caption{Expected counts in the visible gap between the event and the
  lower edge of the observed population. For each family member,
  identified by its shape parameter \(\kappa\), the second column is
  the number of events the member predicts in the interval
  \(269 < N_e < 578\), obtained by folding the parent spectrum of
  Sec.~\ref{sec:leak} through that interval. The third column is the
  Poisson probability of observing zero events given that expectation,
  \(P(0) = e^{-\lambda}\). The interval is defined by the event at
  \(N_e = 269\)~e\(^-\) and the observed population edge at
  \(\log_{10}\mathrm{S2c} = 4.30\), corresponding to
  \(N_e = 578\)~e\(^-\); the lowest point nearby, at
  \(\log_{10}\mathrm{S2c} = 4.25\), lies at
  \(\mathrm{S1c} = 495\)~phd and so outside the comparison region.}
\label{tab:gap}
\end{table}

\section{The electron-recoil contribution to the expected background}%
\label{sec:leak}

The identifiability result of \cref{sec:family} is a statement about
ratios of tail probabilities and is independent of the absolute
normalisation established here. The absolute counts of this section, and
with them the thresholds expressed as fractions of the Letter's quoted
background, are conditional on the parent-spectrum and
selection-efficiency construction set out below.

The S2c ceiling of the WS RoI admits \(N_e<409\)~e\(^-\), and the
\(\mathrm{S1c}\in[500,600]\)~phd comparison region admits
\(N_{ph}\in[4529,5435]\). Since \(N_{ph}=N_q-N_e\) and
\(E=W N_q\) with \(W=13.532\)~eV, the two conditions are jointly
satisfiable only for parents in
\begin{align}
    61.3 < E < 79.1~\mathrm{keV}_\mathrm{ee}\;.
\end{align}
The lower bound is the limit \(N_q\to4529^{+}\), where
\(N_{ph}>4529\) is first satisfiable with \(N_e\geq0\); the upper is
\(N_q\to5844^{-}\), where \(N_e<409\) is last satisfiable with
\(N_{ph}<5435\). The interval \([N_e^\mathrm{lo},N_e^\mathrm{hi}]\) is one
electron wide at both bounds and reaches its full width of 409 electrons
in the middle of the range. The leakage is appreciable only over roughly
\(64\)--\(79\)~\keVee: the scan returns no counts in the column carrying
both cuts at \(63\)~\keVee\ with the trials used there, though it
returns plenty against the ceiling alone, and the support itself ends at
\(79.1\)~\keVee. Integrals below are taken over the kinematic support
\(61.3 < E < 79.1\)~\keVee. Integrating \(P_\mathrm{win}(E)\), the
probability per decay that a parent of energy \(E\) satisfies both cuts
of the comparison region, over the kinematic support gives 
\begin{align} 
	\int P_\mathrm{win}(E)\,\mathrm{d}E = (1.497 \pm 0.090) \times
10^{-6}~\mathrm{keV}. 
\end{align} 
The integral has units of keV because \(P_\mathrm{win}\) is a
dimensionless probability per decay at fixed \(E\). It becomes an
expected count only after folding with the parent spectrum
\(\mathrm{d}N/\mathrm{d}E\), as done below. The parent window lies well
above the 20.02~\keVee\ at which a mean-recombination electron recoil
meets the S2c ceiling. The parent population for this leakage is
therefore drawn from energies three to four times higher than the region
in which the WS RoI contains ordinary electron recoils at all.

The comparison region is a subset of the WS RoI, and the parents that feed
it are the high-energy end of the same continuous population that feeds
the rest of the WS RoI. Below \(61.3\)~\keVee, \(N_q <
4529\) and no fluctuation can raise \(N_{ph}\) into the S1c window;
above \(79.1\)~\keVee, keeping \(N_e < 409\) forces \(N_{ph} >
5435\), outside the window. Table~I of the Letter quotes the total RoI counts, which are
dominated by low-energy parents; converting them into an expectation for
the comparison region therefore requires the ratio of the parent
spectral density in the window to the average density over the RoI
acceptance, which is the window-to-RoI ratio of the spectral densities at the two intervals. For the
\(\beta\)-decay components this ratio depends on the shape of the
\(\beta\) spectrum.

We take the internal-\(\beta\) composition from the collaboration's own
background determination~\cite{Bkg}, whose Table~VI gives counts in the
SR1 WIMP search region \(^{214}\)Pb\(\,:\,^{212}\)Pb\(\,:\,^{85}\)Kr
=164:18:32, and apply those \emph{fractions} to the 1341
internal-\(\beta\) events of Table~I of the Letter. The counts
themselves are not transported: Ref.~\cite{Bkg} reports the first
science run at 193~V/cm over 60 live days, whereas the Letter reports
220 live days, at the \(96.5\)~V/cm of the released
configuration~\cite{Recomb}; the Letter states no drift field for this
exposure in its text, though Table~S3 of the Supplement quotes the NEST
defaults it departs from ``at 97 V/cm''. The absolute rates therefore differ while the
relative composition of the radon chains and \(^{85}\)Kr is expected to
be stable. The remaining components are taken directly from Table~I of
the Letter; ``\(\nu\)~ER'' denotes elastic scattering of solar neutrinos
on atomic electrons, whose recoil spectrum is flat to better than a
percent across both intervals. The digitised spectra used here are
provided in the archive~\cite{Archive}.

The six rows below carry \(1646.9\) of the \(1706.8\) expected counts of
Table~I of the Letter. That total is the sum of the expected column, on
which the thirteen rows of the table close exactly; the \(1713\pm39\)
printed at the foot is the post-fit total, which the fit raises above
the expectation, so it is not the figure a reconstruction of the
expectation should be compared with. The \(59.9\) counts this
reconstruction omits are likewise all accounted for in the table. Of them,
\(0.2\) are the three entries that are not \(\beta\) continuum: atmospheric
and \(^{8}\)B neutrinos, which are nuclear recoils, and MSSI, which is a pair
of electron recoils sharing a single ionisation signal and so appears
nuclear-recoil-like without being a nuclear recoil. None is part of the parent
population we reconstruct. The Letter treats the MSSI interpretation of this
event at length. It reconstructs the two deposits such an event would require,
notes that a first scatter of only \(\sim\!12\)~keV more than \(20\)~cm from
the boundaries of the TPC is unlikely on \(\gamma\)-ray attenuation grounds,
finds the event's pulse shape less consistent with a reverse-field-region MSSI
than with a recoil at the reconstructed position, and reports that its
\(\mathrm{S1}\) hit pattern cannot separate a single scatter there from a wall
MSSI; the category is otherwise held down by its rate, \((4.9\pm4.9)\times
10^{-3}\) counts in Table~I. That is a distinct mechanism from the continuum
leakage considered here, and we take no view on it. The other \(59.7\) are electron-recoil components. Three
are line sources outside the window or at its edge: \(^{83m}\)Kr
(\(17.1\) counts, at \(41.5\)~keV) and \(^{127}\)Xe\,\(+\,^{125}\)Xe
(\(1.5\)) lie below it, and accidental coincidences (\(2.7\)) are not a
parent spectrum at all but a pairing of unrelated \(\mathrm{S1}\) and
\(\mathrm{S2}\) pulses, which the present construction cannot treat.
Detector ERs (\(8.5\)) are a \(\gamma\) continuum whose shape in the
window we have no published spectrum for; at half a per cent of the
total they are inside the uncertainty assigned below. The remaining two
are \(^{124}\)Xe (\(21.0\)) and \(^{125}\)I (\(8.9\)), whose
double-vacancy modes release \(64.3\) and \(67.3\)~keV and therefore
fall \emph{inside} the parent window. These are precisely the components
for which the Letter records an unpropagated tail-projection systematic,
and their charge yield is suppressed relative to a \(\beta\) decay of
the same energy by an amount that Ref.~\cite{Recomb} measures for
single-vacancy captures and for the \(^{124}\)Xe LL mode but not for the
KK mode at \(64.3\)~keV. Because the leakage of a monoenergetic line
depends on that suppression through the exponential tail of the
response, a quantitative estimate would rest entirely on an
extrapolation. We therefore exclude them and record that
\(N^\mathrm{ER}_\mathrm{leak}\) below is the continuum contribution
only.

One row needs a caveat of its own. Table~I of the Letter combines
\(\mathrm{CH_3T}\) and \(^{14}\)C in a single entry of \(55.3\) counts
and does not give the split. Tritium has an endpoint of \(18.6\)~keV and
contributes nothing whatever to a window at \(61\)--\(79\)~\keVee, so
treating the entry as \(^{14}\)C throughout is the conservative choice:
if the whole \(55.3\) were tritium instead, the window rate would fall
from \(123.2\) to \(121.2\) per \keVee, a change of \(1.7\%\). The
ambiguity is bounded by that and is far inside the twenty per cent
assigned below.

\begin{table}
\begin{center}
\begin{tabular}{lrrrr}
\toprule
Component & RoI counts & \(\mathrm{d}N/\mathrm{d}E\) (RoI) & window/RoI & \(\mathrm{d}N/\mathrm{d}E\) (window)\\
\midrule
\(^{214}\)Pb (\(Q=\)\SI{1018}{keV})\(^{\dagger}\) & 1027.7 & 52.97 & 1.057 & 56.00\\
\(^{212}\)Pb (\(Q=\)\SI{569.1}{keV})  &  112.8 &  5.81 & 0.973 &  5.66\\
\(^{85}\)Kr  (\(Q=\)\SI{687}{keV})  &  200.5 & 10.33 & 0.949 &  9.82\\
\(^{136}\)Xe \(2\nu\beta\beta\) &  110.0 &  5.67 & 7.491 & 42.47\\
\(^{14}\)C   (\(Q=\)\SI{156}{keV})  &   55.3 &  2.85 & 0.720 &  2.05\\
\(\nu\) ER (see text)       &  140.6 &  7.25 & 1.000 &  7.25\\ \midrule
Total (window)            &      &      &      & 123.2\\ \bottomrule
\end{tabular} 
\end{center} 
\noindent\(^{\dagger}\)\,We use the ENSDF ground-state
\(Q\)-value~\cite{ENSDF}; evaluations differ by a few keV, which is
immaterial for a ratio taken far below the endpoint.
\caption{The parent spectrum over the window, by component. The six
  rows are the components that contribute to the continuum
  electron-recoil background; the \(^{124}\)Xe and \(^{125}\)I
  double-vacancy lines that also fall in the window are discussed in
  the text and excluded here. The first column gives the component and
  its \(\beta\)-decay \(Q\)-value; the second gives the number of
  expected events in the WS RoI, as tabulated in Table~I of the
  Letter~(for \(^{214}\)Pb, \(^{212}\)Pb and \(^{85}\)Kr, the entries
  are obtained by applying the composition of Ref.~\cite{Bkg} to the
  combined internal-\(\beta\) total). The third column is the RoI count
  divided by the \(19.40\)~keV\(_{ee}\) acceptance width; the fourth
  is the ratio of the spectral density in the window to the average
  over that acceptance; and the fifth is the product of the third and
  fourth. The Total entry applies to the final column only.
  Units are counts per keV\(_{ee}\) for the two \(\mathrm{d}N/\mathrm{d}E\)
  columns and dimensionless for the window/RoI column.}
\label{tab:parent}
\end{table}

\noindent The \(\mathrm{d}N/\mathrm{d}E\) columns are in counts per
\keVee; the window/RoI column is dimensionless; the
Total entry applies to the final column only. The total, \(\simeq 123.2\) per
\keVee, is the parent rate over the window that enters
the folding. The three components whose spectra are nearly flat across
the relevant range, \(^{214}\)Pb, \(^{212}\)Pb and \(^{85}\)Kr, have
window-to-RoI ratios within six percent of unity. Their shapes are taken
from the calculation of Ref.~\cite{BetaShapes}, which is the treatment
used in the collaboration's own background model~\cite{Bkg}: the
\(g_\mathrm{A}=0.85\) curves including atomic exchange, for the
ground-state branches, which are the only ones that reach the RoI
without an accompanying \(\gamma\) ray. Exchange raises the spectra
below \(\sim\)\SI{10}{keV}, which lies inside the RoI denominator and
therefore lowers each ratio relative to an allowed-shape treatment by
3.9\%, 1.7\% and 10.5\% respectively, the largest shift being for the
first-forbidden unique transition of \(^{85}\)Kr. Propagating the
published uncertainty bands moves the \(^{214}\)Pb and \(^{212}\)Pb
ratios by \(\pm\)1--2\%, since the bands are strongly correlated between
the two intervals and largely cancel in a ratio. 

\(^{14}\)C has \(Q=\)\SI{156}{keV}, so the
parent window lies at 40--50\% of its endpoint where the spectrum is
already falling, and its ratio is 0.72. The \(^{136}\)Xe component is
not a \(\beta\) decay but a two-neutrino double-\(\beta\) spectrum peaking near
\SI{700}{keV}; its ratio is 7.49 and it supplies a third of the total
window rate from \(110\) events, under seven per cent of the \(1647\)
counts the six rows below carry\@. It is listed separately there and is not
part of the \(1341\) internal-\(\beta\) events, which the
\(^{214}\)Pb\,:\,\(^{212}\)Pb\,:\,\(^{85}\)Kr split of
\(164{:}18{:}32\) exhausts exactly. We evaluate its summed-electron
spectrum in the standard form~\cite{KotilaIachello}, \((Q-T)^5\) times the
convolution of \(F(Z,\epsilon)p\epsilon\) over the two electrons with
\(Z=56\); the Primakoff-Rosen~\cite{Primakoff:1959chj} closed form would give \(7.88\) in place of
7.49, a 5\% difference that moves the total by 1.8\%. Atomic
exchange has not been calculated for this isotope; it has been computed
for \(^{100}\)Mo~\cite{Nitescu}, where it produces a steep low-energy rise
in the single-electron spectrum of the kind seen in the \(\beta\)
calculations above. Since it would enhance both of our intervals the
effect largely cancels in the ratio: imposing a \(\beta\)-like
per-electron enhancement of 10\% at threshold lowers the \(^{136}\)Xe
ratio by 6\% and the total by 2\%, and even a 30\% enhancement
keeps the shift on the total below 6\%. This is the largest single
spectral uncertainty in the table and is contained within the 20\% we
assign to \(\mathrm{d}N/\mathrm{d}E\). RoI rates are counts divided by the
width of the electron-recoil acceptance of the WS RoI, namely
20.02--0.62=\(19.40\)~\keVee. The upper bound
\(20.02\)~\keVee is the energy at which a
mean-recombination electron recoil meets the S2c ceiling; the lower
bound \(0.620\)~\keVee is the energy at which such a
recoil first satisfies the S1c and S2c cuts. The two are not obtained
the same way, and the floor is the more delicate of the two.

The ceiling follows from
solving \(Q_y(E)\,E=409\) with the tuned charge-yield model. The floor
cannot: below about \(1\)~\keVee\ that parametrisation returns more
electrons than there are ions, exceeding the no-recombination ceiling
\(Q_y^{\max}=1/[W(1+N_{ex}/N_i)]=62.6\)~e\(^-\)/keV, and in any case the
WS RoI is three cuts rather than one. Taking \(\mathrm{S1c}>3\)~phd and
\(\mathrm{S2}>645\)~phd together gives \(N_q>45.9\) and
\(0.62\)~\keVee, which we adopt. It is a boundary of the acceptance
rather than a threshold events cross: sub-keV electron recoils enter only
on a recombination fraction well above the value the capped model
returns. Treating the floor through the S2c cut alone, with or without
the cap, instead gives \(0.19\) or \(0.26\)~\keVee; the three
treatments span \(2.2\%\) in the acceptance width, which is contained in
the \(20\%\) we assign to \(\mathrm{d}N/\mathrm{d}E\). The derivation is
given in \texttt{notes/acceptance-floor.md} of the archive.

There is one assumption that cannot be tested externally. Ratios of counts to rates
presume that the electron-recoil selection efficiency is the same in the
RoI interval 0.62--\(20.02\)~\keVee\ as in the parent window
61--\(79\)~\keVee. The efficiency against energy for this analysis is
not published, and if it falls at low energy the RoI denominator is
overestimated and every ratio in the table rises together. The
associated data release~\cite{LZdata} comprises the
effective-field-theory limit curves and does not include the selection
efficiency or the event-level data in
\(\{\mathrm{S1c},\log_{10}\mathrm{S2c}\}\). Neither was released with the
WS2024 analysis~\cite{WS2024} either: what that record supplies is a
\textsc{nest} v2.4.0 header file reproducing its response models, its
published efficiency is a nuclear-recoil signal efficiency rather than an
electron-recoil selection efficiency, and its region of interest ends at
\(\mathrm{S1c}=80\)~phd, far below the region at issue here. The precedent
that release does set is for publishing the configuration itself, which is
what item~4 below asks for. Since the effect is common to
all components it does not cancel, and we regard it as the leading
uncertainty on \(\mathrm{d}N/\mathrm{d}E\), ahead of any spectral-shape
effect discussed above. One such effect deserves naming, since the
\(\beta\) shapes are the input we take from outside the Letter: the
calculations of Ref.~\cite{BetaShapes} are quoted over a range of the
weak axial-vector coupling, \(g_A=0.7\)--\(1.0\), across which the
predicted \(^{214}\)Pb rate in a \(1\)--\(15\)~keV interval moves by
\(15\)--\(23\%\). Our use of those spectra is a ratio of densities at
two intervals rather than an absolute rate, so the dependence largely
cancels, but it does not cancel exactly and it is a theory systematic
rather than a measurement one. We assign a 20\% uncertainty.

Folding the two, 
\begin{align}
    \label{eq:Nleak}
    N^\mathrm{ER}_\mathrm{leak}=(1.9\pm0.1_\mathrm{stat}\pm0.4_\mathrm{rate})\times10^{-4}~\text{counts}\;.
\end{align}
The absolute normalisation here affects the value of
\(N^\mathrm{ER}_\mathrm{leak}\) but not the identifiability result of
\cref{sec:family}, which is expressed as a ratio of tail
probabilities at fixed parent energy and is independent of the parent
spectrum entirely. This is the \emph{continuum} electron-recoil component
alone: the \(^{124}\)Xe and \(^{125}\)I double-vacancy lines that fall
inside the parent window are excluded for the reasons given above, so
\cref{eq:Nleak} is a lower bound on the electron-recoil contribution in
that sense as well. The \(0.0106\pm0.0008\)
counts quoted in the Letter cover all components, including accidental
coincidences, multiple-scintillation single-ionisation (MSSI) events and atmospheric
neutrinos, and the electron-recoil contribution computed here sits
comfortably below it.

The scan behind \cref{eq:Nleak} is carried to the quanta level: the region's
requirements are imposed on \(N_{ph}\) and \(N_e\) directly, so the
photon-detection and electron-extraction fluctuations that turn quanta into
\(\mathrm{S1c}\) and \(\mathrm{S2c}\) are absent, and the edges of the
\(\mathrm{S1c}\) window and the \(\mathrm{S2c}\) ceiling are sharp where the
detector makes them soft. We have measured what that costs by repeating the
scan through the detector response and applying the cuts in phd. The parent
support widens from \(64\)--\(79\) to \(60\)--\(85\)~\keVee, and the
window integral rises by a factor of
\begin{align}
    \label{eq:smear}
    \frac{\int P_\mathrm{win}^{\mathrm{S1c,S2c}}\,\mathrm{d}E}
         {\int P_\mathrm{win}^{\mathrm{quanta}}\,\mathrm{d}E}\;&=\;1.17\,,
         \qquad\text{paired count ratio } 1.18\pm0.10\;,
\end{align}
the two columns coming from the same simulated events, so the comparison
carries no sampling error beyond the counts. The quanta-level treatment
therefore \emph{understates} \(N^\mathrm{ER}_\mathrm{leak}\), which becomes
\(2.2\times10^{-4}\) counts, and \cref{eq:Nleak} is conservative in the
direction that matters. The same run reproduces the quanta-level window
integral of the archived scan to four figures, \(1.4968\times10^{-6}\)
against \(1.4970\times10^{-6}\)~keV, which is what licenses the comparison.

\begin{table} \begin{center} \begin{tabular}{lr} \toprule Variation &
	Factor on \(N^\mathrm{ER}_\mathrm{leak}\)\\ \midrule
	\(\sigma_{N_e}+5\%\) & \(\times4.2\)\\ \(\sigma_{N_e}+10\%\) &
	\(\times14.4\)\\ \(\sigma_{N_e}+15\%\) & \(\times42.0\)\\
\(\sigma_{N_e}+20\%\) & \(\times107\)\\ \(\gamma_{\rm ER}-0.35\) &
\(\times6.3\)\\ \(\gamma_{\rm ER}+0.35\) & \(\times0.11\)\\ \bottomrule
\end{tabular} \end{center} \caption{Factor by which the electron-recoil
	contribution \(N^{\rm ER}_{\rm leak}\) changes under a finite
	variation of the two response parameters to which the leakage is
	most sensitive. The first column gives the variation: a
	fractional increase in the response width \(\sigma_{N_e}\), or
	an additive change in the event-level skewness \(\gamma_{\rm
	ER}\) from its default value of \(1.892\). The second column is
	the resulting factor on the per-decay probability, evaluated at
	the event-conditioned energy with all other response quantities
	held fixed. The response is strongly non-linear in the size of
	the perturbation, so the entries do not scale simply with the
	size of the variation; the effective logarithmic derivative
	falls from \(31.0\) at zero variation to \(20.9\) at \(+20\%\),
	and the table should be read in place of extrapolating
Eq.~\eqref{eq:sensitivities}. Only upward width variations are
tabulated: downward variations reduce the leakage.}
\label{tab:sensitivity} 
\end{table} 

The effect on the ratios of \cref{sec:family} cannot be measured the
same way, the scan existing only for the response the configuration
implements. Modelling the two fluctuations and calibrating the model
against \cref{eq:smear}, the ratios are compressed by \(7\%\) at
\(\kappa=1.8\) rising to \(13\%\) at \(\kappa=1.4\). The compression and
the rise in the absolute leakage act in opposite senses on the
thresholds below and very nearly cancel: each \(\kappa\) there moves by
under \(0.01\), so the thresholds are unchanged at the precision we
quote them. We retain the quanta-level numbers throughout, as the
conservative choice, and the measured comparison is in the archive.

The expectation of \cref{eq:Nleak} rests on a per-decay probability
\(P\simeq1.3\times10^{-7}\), a displacement of
\(\Delta/\sigma_{N_e}\simeq4.1\) from the mean of the response. At that
displacement the leakage responds steeply to the response parameters:
\begin{align}
    \label{eq:sensitivities}
    \frac{\partial\ln P}{\partial\ln\sigma_{N_e}}=31.0\;,\qquad
    \frac{\partial\ln P}{\partial\gamma_\mathrm{ER}}=-5.8\;,
\end{align}
both approximately constant in energy across
45--\(110\)~\keVee\ (\cref{fig:f2}, right panel).
The response is strongly non-linear in the size of the perturbation.
Write \(\epsilon\) for a fractional variation in \(\sigma_{N_e}\), so
that \(\sigma_{N_e}\to\sigma_{N_e}(1+\epsilon)\), and let
\(\mathcal{R}(\epsilon)=P(\sigma_{N_e}(1+\epsilon))/P(\sigma_{N_e})\)
be the resulting factor on the leakage probability. The instantaneous
logarithmic derivative \(\partial\ln P/\partial\ln\sigma_{N_e}\) is
\(31.0\) at the reference point \(\epsilon=0\), but falls to \(20.9\)
at \(\epsilon=0.20\): as the width grows the S2c ceiling at
\(N_e=409\) sits relatively less far from the mean of the
distribution, and the sensitivity to further variation drops. A finite
variation is therefore governed not by the derivative at the reference
point but by its average over the interval, the chord
\begin{align}
    \bar{c}(\epsilon)\;=\;\frac{\ln \mathcal{R}(\epsilon)}{\ln(1+\epsilon)}\;,
\end{align}
which over \([0,0.20]\) is \(25.6\). The factor for a \(20\%\)
increase is accordingly \(\mathcal{R}=(1.20)^{25.6}\approx107\), not
the \(\exp(31.0\times0.20)\approx490\) that a naive extrapolation of
the local derivative would give. Finite variations should therefore be
read from the table below; \cref{eq:sensitivities} is accurate only
for variations small enough that the fall to \(20.9\) across the
interval is negligible.

The response to \(\gamma_\mathrm{ER}\) is asymmetric, and only upward
width variations are tabulated in \cref{tab:sensitivity}.

For a given target expectation we
solve \(N^{\mathrm{ER}}_{\mathrm{leak}}(\kappa)=N^{\mathrm{ER}}_{\mathrm{leak}}\,
R(\kappa)\) for the shape parameter \(\kappa\) of \cref{eq:family}, where
\(R(\kappa)\) is the tail-probability ratio of \cref{sec:family} and
\(N^{\mathrm{ER}}_{\mathrm{leak}}=1.84\times10^{-4}\) counts the unrounded nominal
value of \cref{eq:Nleak}; the quantile deviation quoted alongside is then
evaluated at that \(\kappa\). This uses the ratio at the event-conditioned
energy, which \cref{sec:family} shows differs from the
window-weighted value by at most two percent. The thresholds are

\begin{table}
\begin{center}
\begin{tabular}{lrrr}
\toprule
\(N^{\mathrm{ER}}_{\mathrm{leak}}\) reaches & \(\kappa\) & quantile deviation & interpretation\\
\midrule
\(1\%\) of the quoted total   & 2.04 & \(0.04\%\) & \emph{lighter} than the nominal tail\\
\(10\%\) of the quoted total  & 1.87 & \(0.14\%\) & ceases to be negligible\\
the quoted total              & 1.66 & \(0.47\%\) & ER alone reaches the quoted total\\
\(0.1\) counts                & 1.39 & \(1.10\%\) & an appreciable fraction of one event\\
\bottomrule
\end{tabular}
\end{center}
\caption{The shape parameter \(\kappa\) at which the electron-recoil
  contribution reaches specified fractions of the quoted background.
  For each target in the first column, the second column gives the
  value of \(\kappa\) in the family of Eq.~\eqref{eq:family} at which
  \(N^{\rm ER}_{\rm leak}\) equals that target, obtained by solving
  \(N^{\rm ER}_{\rm leak}(\kappa) = R(\kappa)\,N^{\rm ER}_{\rm leak}\)
  continuously in \(\kappa\) rather than interpolating the family table;
  the third column gives the corresponding quantile deviation from the
  executable model's central characterisation. The nominal expectation
  is \(1.74\%\) of the quoted total, so the first row lies below the
  nominal and is reached only by a tail \emph{lighter} than the
  executable model's; the rows below it are the ones at issue. The
  final column is an interpretive gloss, not a derived quantity.}
\label{tab:thresholds}
\end{table}

\noindent The nominal expectation is already \(1.74\%\) of the quoted
total, so the first row of the table lies below it and is reached only by
a tail \emph{lighter} than the executable model's; the rows below it are
the ones at issue. The ladder is short: a shape agreeing with the central
characterisation to within a percent in the quantiles spans the whole of
it, from an electron-recoil component that is a tenth of the quoted
background to one that is an appreciable fraction of a single event. None
of these deviations is excluded by anything published, and all lie inside
the range over which \cref{sec:family} shows the absence of intermediate
events to be uninformative.

What we cannot do is carry this through to the reported significance. The
\(2.6\sigma\) the Letter quotes is a global significance, obtained by a
look-elsewhere correction over a model space of \(616\) hypotheses reduced
to \(293\) distinguishable spectra; the profile likelihood ratio itself
returns the local significances tabulated model by model in Tables~S7
and~S8 of the Supplement, whose maximum is \(3.4\sigma\). The ER
component enters as one term among several in one region of a
two-dimensional unbinned likelihood, which Eq.~3 of the Supplement writes
out as a product over the science, prompt-veto and delayed-veto samples
with their rates tied by tagging efficiencies. What is unpublished is that
likelihood's content rather than its form: the background densities, the
fitted parameter values and the coordinates of the other events in the
samples are none of them available---the event of interest's own
\(\mathrm{S1c}\) and \(\mathrm{S2c}\) are published, and are what
\cref{eq:e1} inverts--- so converting a change in
\(N^{\mathrm{ER}}_{\mathrm{leak}}\) into a change in the test statistic
is not something we can do, and we do not attempt it. The table above is a statement about the background
expectation for the comparison region, and about where the electron-recoil
component stops being a small correction to it.

\section{What would determine the response tail}
\label{sec:settle}

What would settle the question this paper raises is a measurement of the response
distribution. The families of \cref{sec:family} are separated by their
shape at a displacement of \(4.1\) standard deviations, not by their
first three moments, which they share; the quantity that discriminates
among them is therefore the low-side tail of the \(N_e\) distribution
itself. A measurement at \(N_q\simeq5\times10^{3}\) reaching the
\(10^{-5}\) level would substantially constrain the shapes admissible
in \cref{sec:family}, and direct validation of the response at the
\(10^{-7}\) level would test the shape at the displacement where the
inference is made. No improvement in the width measurement, by itself,
resolves the between-family ambiguity. A width measurement does move the
leakage, and we quantify how much below; but the families are
constructed with \(\sigma_{N_e}\) held at the value the executable model
returns, so the spread between them lies entirely in the shape at
\(4.1\sigma\), and tightening the width leaves that spread where it is.
Width precision and tail-shape identifiability are distinct experimental
requirements.

It is fair to ask whether the tail has to be reached at all. The
families differ in the bulk as well as in the tail: the density
deviation column of \cref{sec:family} runs from three to eighteen per
cent, so a shape fit to a \(\beta\) calibration of ordinary size might
separate them without ever reaching \(10^{-5}\). We have evaluated what
that would take, in the form a real calibration would be obliged to use.
Take the heavier member as the truth \(q\), fit the implemented response
to it with its mean, width, and skewness free, and read the separation
from the Asimov expression

\begin{align}
Z^{2} \;=\; 2N\,D\!\left(q\,\|\,\hat{p}\right),
\qquad
D(q\,\|\,p)=\int q\,\ln\!\frac{q}{p}\,\mathrm{d}N_e ,
\label{eq:asimov}
\end{align}

\noindent where \(D(q\,\|\,p)\) is the Kullback--Leibler
divergence~\cite{Kullback:1951}. The expression is the asymptotic
non-centrality of the likelihood-ratio statistic for \(N\) events drawn
from \(q\) and fitted to \(\hat p\)~\cite{Wilks,Asimov}, evaluated on
the Asimov dataset in the sense of Ref.~\cite{Asimov}, with \(N\) the
number of calibration events in a slice at fixed \(N_q\) and \(\hat{p}\)
the implemented response at its best-fitting moments. The moments are
free because a calibration determines them from the same events that
must supply the shape; the members share them exactly by construction,
so the fit begins at the truth, but it is not obliged to stay there.

The prescription is the following, and the code is in the
archive~\cite{Archive}. Both \(q\) and \(\hat p\) are continuous
densities in \(N_e\) on \(0<N_e<8000\)~e\(^-\), wide enough that both
tails of every member fall below double precision inside it, and are
renormalised there. \(\hat p\) is the skew-normal of
\cref{eq:skewnormal} reparameterised by its first three moments, which
are varied as multiplicative factors on the reference mean, width and
standardised skewness; the skewness is bounded by the skew-normal's own
limit \(|\gamma_1|<0.9953\), and we record whether a fit reaches it, so
that a boundary is not read as an interior optimum. The divergence is
minimised over those three factors by the simplex method of Nelder and
Mead~\cite{NelderMead} as implemented in SciPy~\cite{SciPy}, which is the
profiling: no further nuisance enters. Where the measured charge
resolution is folded in, it is applied to \(q\) and to \(\hat p\) alike,
as an \(N_e\)-dependent Gaussian kernel, before the divergence is taken.
The bulk-only variant restricts the integral to the central
\(1-2\times10^{-4}\) of \(q\) by its own distribution function and does
not renormalise, so that it measures the separating power available
inside that range rather than the separation of two truncated
distributions. As a check that the calculation resolves what it reports,
evaluating it at \(\kappa=2\)---where the member \emph{is} the
implemented response and the divergence must vanish---returns
\(2\times10^{-10}\), four orders below anything in the table, and the
answers agree to four figures on grids of \(4\), \(2\) and
\(1\)~e\(^-\).

With the measured charge resolution folded in, the requirement is

\begin{table}
\begin{center}
\begin{tabular}{lrrrr}
\toprule
member of the ladder above & \(\kappa\) & \(3\sigma\) & \(5\sigma\) & bulk only, \(3\sigma\)\\
\midrule
reaches a tenth of the quoted total & 1.87 & \(2.9\times10^{4}\) & \(8.0\times10^{4}\) & \(7.6\times10^{4}\)\\
reaches the quoted total            & 1.66 & \(3.2\times10^{3}\) & \(8.8\times10^{3}\) & \(4.1\times10^{3}\)\\
reaches \(0.1\) counts              & 1.39 & \(5.4\times10^{2}\) & \(1.5\times10^{3}\) & \(6.0\times10^{2}\)\\
\bottomrule
\end{tabular}
\end{center}
\caption{Number of calibration events in a slice at fixed \(N_q\)
  needed to separate an alternative response shape from the
  implemented one, evaluated on the Asimov dataset from
  Eq.~\eqref{eq:asimov}. Each row is one member of the ladder of
  Table~\ref{tab:thresholds}, identified by the shape parameter
  \(\kappa\) at which that member lifts the electron-recoil
  contribution to the stated level. The columns give the requirement
  at three standard deviations, at five standard deviations, and at
  three standard deviations when the divergence is restricted to the
  central \(99.98\%\) of the truth (the ``bulk only'' column). The
  requirement scales as \(Z^2\) by construction, so the \(5\sigma\)
  column is a factor of \((5/3)^2 \simeq 2.8\) larger than the
  \(3\sigma\) column in each row. All entries have the measured charge
  resolution folded in; without it they would be lower by a factor of
  about \(1.07\). The calculation is described in the text, and the
  code is in the archive~\cite{Archive}.}
\label{tab:asimov}
\end{table}

\noindent Three features of that calculation matter more than the
numbers themselves. The nuisances are cheap: holding the three moments
fixed instead of fitting them lowers the requirement by only ten to
twenty per cent, because moments the two distributions share exactly
cannot absorb a difference of shape. What the fit does instead is move
the skewness, and in every row by about one standard error of the
skewness at that same \(N\). The requirement is also insensitive to the
detector: folding in the charge resolution raises every entry by seven
per cent, so this is a statement about the response and not about the
readout.

The third feature is that how much of the separation comes from the tail
is not a property of the method but of the shape being tested, and it
varies across the ladder. Restricting the divergence to the central
\(99.98\%\) of the truth---discarding the outer \(10^{-4}\) at each end,
which a calibration of this size populates only sparsely, some six
events at \(N=3\times10^{4}\) and well under one at
\(N=5\times10^{2}\)---leaves \(38\%\) of the separating power at
\(\kappa=1.87\), \(78\%\) at \(\kappa=1.66\) and \(91\%\) at
\(\kappa=1.39\). The last column of the table is the corresponding
requirement. So for the heavier members the separation is substantially
a bulk measurement: excluding the outer tails raises the requirement by
only \(28\%\) at \(\kappa=1.66\) and \(11\%\) at \(\kappa=1.39\). It is
the member nearest the implemented response, the one that lifts the
electron-recoil component to a tenth of the quoted background, for which
the tails carry most of the information, and that member is also the
most expensive: \(3\times10^{4}\) events at three standard deviations,
\(8\times10^{4}\) at five, and \(8\times10^{4}\) at three with the outer
tails discarded.

A shape fit is therefore an alternative route for part of the ladder and
not for all of it. Where the alternative tail is heavy enough to bring
the electron-recoil component up to the quoted background, a few
thousand events in a slice would separate it, and mostly from the bulk;
where it is only heavy enough to bring that component to a tenth of the
quoted background, the requirement rises by an order of magnitude and
moves into the same tail that item~1 below asks to be measured directly.
A \(\beta\) calibration is a broad spectrum, and what fraction of such a
sample falls in one slice is not something the published material lets
us compute, so these are scaling statements and not limits.

There is precedent for a characterisation of this kind, though not for
the measurement itself. Facing the same difficulty---a discrimination
leakage required far below what direct counting can reach---the
DEAP-3600 collaboration published a full model of its
discrimination-parameter distribution, together with leakage
probabilities as a function of energy and nuclear-recoil acceptance and
the systematic variations that change those predictions~\cite{DEAP}.
The published predictions extend well below the level at which direct
event counting could determine the leakage, and they are accompanied by
the model and its uncertainties rather than by a single number. That
last clause is the one that matters here. A model
extrapolated into the tail is what the released configuration already
provides; what DEAP-3600 supplied in addition, and what is missing in
the present case, is a stated uncertainty on the shape at the
displacement where the extrapolation is used. The physics is not
analogous, DEAP-3600 is single-phase argon, and its discriminant is a
scintillation-timing ratio rather than a charge fraction. We are
not aware of a closer precedent in a xenon experiment. Item~1 below asks
for the stronger form of the same deliverable: the tail measured rather
than modelled, at the level the inference requires.

It is worth setting out what accuracy \emph{is} published for the
\(\beta\)-calibrated response, because it is the central characterisation
and nothing else. Ref.~\cite{WS2024} states that the tuned model
reproduces the ER band median and width to better than \(0.2\%\) and
\(5\%\) in \(\log_{10}\mathrm{S2c}\) respectively. Both are properties of
the centre of the response, and both are measured on tritium, whose
endpoint is \(18.6\)~keV. We are not aware of a published \emph{measured}
electron-recoil leakage below the nuclear-recoil median, at any energy,
against which the tail could be checked; the Letter reports leakage for
\(\mathrm{S1c}>250\)~phd as predicted rather than measured.

Two features of that calibration bear directly on the present case. The
first is its reach in \(\mathrm{S1c}\): the tritium and \(^{14}\)C data
are cut off above \(\mathrm{S1c}=500\)~phd to avoid \(^{133}\)Xe
contamination from a preceding neutron calibration~\cite{LZ:2026axp}, and
the event sits at \(540.1\)~phd. The comparison region is therefore
entirely above the \(\mathrm{S1c}\) range over which the methane-injected
\(\beta\) sources constrain the response, leaving \(^{212}\)Pb as the
\(\beta\) calibration that reaches it. The second is the width. The
published agreement is \(5\%\), and \cref{eq:5b} inverted below gives
\(+6.2\%\) as the width excess at which the electron-recoil contribution
reaches a tenth of the quoted background: the two are the same size. The
comparison is indicative rather than exact, the \(5\%\) being quoted for
the earlier response model, which the Supplement of the Letter records as
having failed to reproduce the measured band widths across the calibration
range and which the two-component fluctuation model replaced. The Letter
reports that the replacement agrees with those widths but gives no
corresponding numerical figure. A width known to something like \(5\%\)
at the energies where tritium populates the band is therefore close to,
but not inside, the accuracy this inference would need. It is not established at the \(\mathrm{S1c}\) of the event.

A width measurement, separately, addresses only the sensitivity of the
leakage within a single family, and not the spread between families that
is the identifiability result. Inverting \cref{eq:5b} numerically for
the width at which the electron-recoil contribution reaches a tenth of
the quoted background, with the shape held at \(\kappa=2\), gives
\(+6.2\%\): a
\(\sigma_{N_e}\) known to that accuracy would keep the contribution
below a tenth of the quoted total, and \(+16.6\%\) is where it reaches
the total itself. Both are far weaker requirements than the tail
measurement above, and neither touches the between-family result.

Four items, one measurement, two tabulations and one clarification,
would determine the quantities this paper identifies, in decreasing
order of how directly they bear on the inference:

\begin{enumerate}
\item The low-side tail of the \(N_e\) distribution at
    \(N_q\simeq5\times10^{3}\), to the \(10^{-5}\) level and with its
    uncertainty, which would constrain the shape and not only the width.
    It must reach above \(\mathrm{S1c}=500\)~phd, where the
    methane-injected \(\beta\) data are presently cut off and the
    comparison region begins.
\item The event-level search data of this analysis in
    \(\{\mathrm{S1c},\log_{10}\mathrm{S2c}\}\), tabulated, which would make
    the intermediate-events test of \cref{sec:family} reproducible from
    published information rather than from the single population edge we
    read off Fig.~4 of the Letter.
\item The electron-recoil selection efficiency against energy, which
    would bound the leading uncertainty on \(\mathrm{d}N/\mathrm{d}E\)
    (\cref{sec:leak}) and with it the absolute rates quoted there; this
    needs no new detector measurement if the corresponding internal
    table can be released.
\item The value of \(\gamma_{\rm ER}\), the \texttt{OldW13eV}
    convention, and the normalisation convention applied to Eq.~4,
    stated alongside the response tables, which would let the printed
    specification be evaluated to the same numbers as
    \texttt{RecombOmegaER} and would distinguish \(\gamma_{\rm ER}\)
    from the \(\alpha_i\) of that equation.
\end{enumerate}

\noindent None of these necessarily requires a new detector or a new
exposure. Items~2 and~3 could be supplied from information already in
hand, item~4 is a specification, and item~1 would extend an existing
calibration programme into the relevant tail.

\section{Summary}

The published characterisation of the LZ electron-recoil response fixes
the centre of that response and does not determine the probability on
which the background expectation for the event's region depends; nor
does it state how far the tail that carries that probability may depart
from the released model's. Across the families constructed here,
distributions that reproduce the released model's mean, variance, and
standardised skewness exactly, and its event-conditioned quantiles to
about one per cent, differ in that probability by more than five orders
of magnitude. Three families agree on this within a factor of about two
at matched quantile deviation, and span two distinct construction
mechanisms; the third does not contain the executable model as a member,
which makes it unlikely that the result is an artefact of any one
parameterisation of the deformation.

The thresholds are close. Solving continuously in the shape parameter, a
tail agreeing with the central characterisation to \(0.14\%\) in the
quantiles lifts the electron-recoil contribution above a tenth of the
quoted background; \(0.47\%\) brings it level with that background
entire; \(1.10\%\) brings it to \(0.1\) counts. None of them is excluded
by the published information examined here. These figures are for the
probability integrated to the \(\mathrm{S2c}\) ceiling; the density at
the event's own charge, which is what an unbinned likelihood weighs,
separates the same families by a further order of magnitude. The
identifiability result itself does not depend on how the absolute rate
is normalised; the thresholds, expressed as fractions of the quoted
background, do, and they carry the parent-spectrum and
selection-efficiency construction of \cref{sec:leak} with its twenty per
cent uncertainty. Carrying the simulation only to the quanta level
understates the absolute expectation by a measured factor of \(1.17\)
and leaves those thresholds where they are.

The LZ publication's own empirical counter-argument does not separate
them. Within \(500<\mathrm{S1c}<600\)~phd the observed population
extends down to \(N_e=578\)~e\(^-\), so the interval observed to be
empty is \(269<N_e<578\). Every family we construct predicts well under
one event there, and none is appreciably constrained by the observed
absence of events in that interval.

The situation is not peculiar to this event. Any rare-event search that
interprets a small number of candidates against an expectation computed from
the far tail of a calibrated response faces the same question, and the
published summary of a response need not determine such a tail.

What is missing is a measurement. The low-side response tail at
\(N_q\simeq5\times10^{3}\), measured to the \(10^{-5}\) level, would
constrain the shape and not merely the width;  and the event-level
search data of this analysis in \(\{S1c,\log_{10}S2c\}\), tabulated,
would make the test of \cref{sec:family} reproducible from published
information; the earlier WS2024 release does not supply it, its region
of interest ending at \(\mathrm{S1c}=80\)~phd. The second is a
tabulation of measurements already made; the first extends into the
relevant tail a response characterisation of the kind another
collaboration has already published, and neither needs a new measurement
campaign.

\appendix

\section{Implementation details}%
\label{app:impl}

The divergences below do not change the leakage probability materially
and are recorded for reproducibility. The one that does: the
normalisation of Eq.~4 of the Supplement, worth a factor of
\(3.2\times10^{3}\), is in \cref{sec:recon}.

One further point is worth recording, not as a divergence in the present
configuration but as a measure of how far the treatment of the response
tail has moved. LZ's first search reports that ``the tritium data are
best modeled with the \textsc{nest} recombination skewness model
disabled'', its supplement adding that the predicted ER and NR band
widths ``had to be reduced to match LZ calibration
data''~\cite{WS2022}. The \textsf{EFT2024} configuration used here
instead has the skewness model enabled, at
\(\gamma_\mathrm{ER}=1.892\) for the event-conditioned energy. The two
analyses are three years, several \textsc{nest} versions and a change of
drift field apart, so this is not an inconsistency and we draw no
conclusion from it. It bears mentioning only because the skewness is the
parameter to which the tail is most sensitive after the width---
\cref{sec:recon} gives \(\partial\ln P/\partial\gamma_\mathrm{ER}=-5.8\)
and a factor of \(57\) across \(\gamma_\mathrm{ER}\in[1.55,2.25]\)---so
a switch of that kind is not a detail of bookkeeping, and which way it
is set has not generally been stated alongside the response tables.

The normalisation referred to in \cref{sec:recon} divides component \(i\) of
\cref{eq:sigmap} by its own value at \(x_i^\mathrm{mode}\), so that the
component takes the value \(A_i\) there:
\begin{align}
    \mathcal{N}_i\;&=\;\Big[e^{-(x_i^\mathrm{mode}-\mu_i)^2/2\sigma_i^2}
        \big(1+\mathrm{erf}[\alpha_i(x_i^{\rm
    mode}-\mu_i)/\sqrt{2}\sigma_i]\big)\Big]^{-1}\;,
\quad
x_i^\mathrm{mode}=\mu_i+\frac{2\alpha_i\sigma_i}{\sqrt{2\pi(1+\alpha_i^{2})}}\;.
\end{align}
The skew-Gaussian mode has no closed form. The expression above is the one
the implementation uses, and it is in fact the skew-Gaussian \emph{mean},
\(\mu_i+\sigma_i\delta_i\sqrt{2/\pi}\) with
\(\delta_i=\alpha_i/\sqrt{1+\alpha_i^2}\), rather than its mode. The
component therefore equals \(A_i\) at its mean and not at its maximum: for
the first the two nearly coincide, \(2.4904\) against a mode at \(2.4920\),
but for the second, whose skew is larger, the mean lies at \(4.1880\)
against a mode at \(4.5235\), and the normalised component reaches
\(1.080\) at its peak. We note this as another point of divergence between
the printed description and the executable model. We adopt the
implementation's expression because it is what reproduces the released
value of \(\omega\); we do not infer whether the naming reflects an
intentional convention or an oversight, the observation being one of
reproducibility rather than of correctness. At the relevant \(N_q\) this
gives \(\mathcal{N}_1=0.963\) and \(\mathcal{N}_2=0.678\). The second
component carries almost all of the effect: it dominates \(\omega\) at this
\(N_q\), and \(1/0.678=1.475\) accounts for the whole of the quoted ratio to
three figures.

The \texttt{OldW13eV} flag of the implementation selects the
\(W\)-value, and through it the quanta-to-energy conversion on which
\(\gamma_\mathrm{ER}\) depends. Its name is slightly misleading: both
branches compute \(W\) from the liquid density rather than fixing it at
\(13.7\)~eV. The implementation evaluates
\begin{align}
    \label{eq:wvalue}
    W_\mathrm{raw}\;&=\;18.7263-1.01\times10^{-23}\,n_e~\mathrm{eV}\;=\;11.550~\mathrm{eV}\,,\quad 
    n_e\;=\;\frac{\rho}{A}N_{\!A}Z=7.105\times10^{23}~\mathrm{cm^{-3}}\,,
\end{align}
at the LZ density \(\rho=2.8687\)~g\,cm\(^{-3}\), and then multiplies by a constant factor
1.17163 when the flag is \texttt{true}:
\begin{center}
\begin{tabular}{lll}
\toprule
\texttt{OldW13eV} & \(W\) at LZ density & Character\\
\midrule
\texttt{true}  & \(13.532\)~eV & \(W_{\rm raw}\) scaled to the value adopted in WIMP searches\\
\texttt{false} & \(11.550\)~eV & \(W_{\rm raw}\) itself, near the \(11.5\)~eV recently measured~\cite{Recomb}\\
\bottomrule
\end{tabular}
\end{center}
\noindent The released configuration sets \texttt{true}, so
\(W=13.532\)~eV, and this is the value used throughout the present
paper, including in the inversion of \cref{eq:e1} and the
acceptance of \cref{sec:leak}. We obtain it both from
\cref{eq:wvalue} and, independently, as \(E/N_q\) from the returned
yields. The flag enters the tail through one channel only. The yields and the
width are evaluated from \(N_q\), which the flag does not touch, so
\texttt{recombProb}, \(\omega\) and \(\sigma_{N_e}\) are numerically
identical between the two settings; the \(W\)-value enters where the
quanta are converted to an energy to evaluate the skewness
parametrisation, and there alone. Switching it therefore moves
\(\gamma_\mathrm{ER}\) from \(1.89\) to \(2.29\) and leaves every other
response quantity where it was, and the leakage falls by a factor of
\(10.3\). Setting \(\gamma_\mathrm{ER}=2.29\) by hand with the flag left
at \texttt{true} reproduces the \texttt{false} result to within the
statistics of the run, which confirms that the convention acts on the
tail through the skewness and not otherwise. The size is what the
logarithmic derivative of \cref{sec:recon} implies:
\(\partial\ln P/\partial\gamma_\mathrm{ER}=-5.8\) over
\(\Delta\gamma_\mathrm{ER}=0.40\) gives a factor of \(10.0\), and the
analytic representation of \cref{eq:5b}, evaluated at the two skewness
values with \(\mu\) and \(\sigma_{N_e}\) held fixed, gives \(12.5\) once
the curvature is included. This is the paper's argument in miniature:
the mean and the width are held exactly, and the tail probability moves
by an order of magnitude under a convention that the printed response
tables do not state.

Table~S3 of the Letter reports \(m_6=0\); the released configuration
contains \(m_6=-1\). In the released implementation this parameter is
never used as a coefficient---the second sigmoid term is hardcoded with
\((0-m_5)\)---and only its sign is read, a negative value activating a
density and \(W\)-value correction applied to the whole ER charge yield.
The effect on the mean yield at these conditions is \(+0.5\%\), which we
include throughout.

\section*{Code and data availability}

All code and derived data needed to reproduce the results of this paper
are archived at Zenodo under
DOI~\textsc{\href{https://zenodo.org/records/23318901}{10.5281/zenodo.23318901}},
with the development repository
at~[\url{https://github.com/triparno/LZ_ER}]. The archive contains the
C++ drivers that instantiate the LZ detector configuration and call the
NEST yield and quanta routines; the parallel scan script and its output,
the \(1.1\times10^{10}\)-event scan of \cref{sec:recon}, which is the
only result here that is not regenerable in minutes; the Python scripts
that construct the distribution families of \cref{sec:family}, perform
the leakage and threshold calculations, and produce every figure; the
digitised \(\beta\) spectra used in \cref{sec:leak}, together with the
extraction scripts, so that the digitisation of the published figures of
Ref.~\cite{BetaShapes} is itself checkable; and the working notes, which
are referred to at several points above and carry calculations omitted
from the text. These are \texttt{notes/acceptance-floor.md}, the
derivation of the low-energy edge of the acceptance;
\texttt{notes/double-vacancy.md}, \(^{124}\)Xe and \(^{125}\)I estimates
for the double-vacancy population discussed in \cref{sec:intro}, which
are not reproduced in the text; and
\texttt{notes/band-reconstruction.md}, which records an attempt to
reconstruct the band of Fig.~2 of the Letter from the released
configuration, the respects in which it did not succeed, and why we
accordingly make no claim resting on it. NEST itself is not
redistributed. Every simulation result was obtained with the public
repository~\cite{NESTcode} at commit \texttt{541b643}, and the archive
records the commit together with build instructions rather than a copy
of the library. Code is released under the MIT licence and tabulated
data under CC~BY~4.0.

\section*{Declaration on the use of generative AI}
During the preparation of this work, the authors used Claude Opus 5
(Anthropic) to assist with manuscript editing, C++ and Python code
development, numerical cross-checking, and preliminary literature
mapping. All AI-generated suggestions, code, and source citations were
fully reviewed and verified by the authors against primary sources. The
authors accept full responsibility for the accuracy and integrity of the
final manuscript.

\bibliographystyle{JHEP}
\bibliography{LZ}

\end{document}